\documentclass[aps,prd,preprint,groupedaddress,nofootinbib,showkeys]{revtex4-1}
\usepackage{amsmath}
\usepackage{graphicx}
\usepackage{lineno}
\usepackage{color}
\newcommand{\hatR}{\mathbf{\hat{R}}}
\newcommand{\xtprim}{(\mathbf{x'},t')}
\newcommand{\xt}{(\mathbf{x},t)}
\newcommand{\rhoprim}{\rho\xtprim}
\newcommand{\Jprim}{\mathbf{J}\xtprim}
\newcommand{\xdiff}{\mathbf{x} - \mathbf{x'}}
\newcommand{\ret}{_\textrm{ret}}
\newcommand{\deltaprim}[1]{\delta^3(\mathbf{x}' - \mathbf{x}_{#1})}
\newcommand{\deltav}{\delta^3(\mathbf{x}' - \mathbf{x}_1 -\mathbf{v}(t'-t_1) ) }
\newcommand{\thetaprim}[1]{\Theta(t' - t_{#1})}
\newcommand{\rect}{\thetaprim{1}-\thetaprim{2}}
\newcommand{\Rect}{\mathrm{\Pi}(t',t_1,t_2)}
\newcommand{\xw}{(\mathbf{x},\omega)}
\newcommand{\xpw}{(\mathbf{x'},\omega)}
\newcommand{\intx}{\int \mathrm{d}^3 x}

\begin{document}

%\preprint{}

\title{Calculations of low-frequency radio emission by cosmic ray-induced particle showers}


\author{Daniel Garc\'\i a-Fern\'{a}ndez, Beno\^\i t Revenu, Didier Charrier, Richard Dallier, Antony Escudie, and Lilian Martin}
%[]{daniel.garcia-fernandez@subatech.in2p3.fr}
%\thanks{}
\affiliation{Subatech, Institut Mines-T\'{e}l\'{e}com Atlantique, CNRS, Universit\'{e} de Nantes, Nantes, France}


\date{\today}

\begin{abstract}
The radio technique for the detection of high energy cosmic rays consists in measuring the electric
field created by the particle showers created inside a medium by the primary cosmic ray.
The electric field is then used to infer the properties of the primary particle. Nowadays,
the radio technique is a standard, well-established technique. While most current experiments
measure the field at frequencies above $20$ MHz, several experiments have reported a
large emission at low frequencies, below $10$ MHz. 
The EXTASIS experiment aims at measuring again and understand this low-frequency electric field.
Since at low frequencies the standard
far-field approximation for the calculation of the electric field does not necessarily hold, in order
to comprehend the low-frequency emission we need to go beyond the far-field approximation.
We present in this work a formula for the electric field created by a particle track 
inside a dielectric medium that is valid for all frequencies. We then implement this formula
in the SELFAS Monte Carlo code and calculate the low-frequency electric field of EAS.
We also introduce the sudden death pulse (SDP), an emission caused by the coherent
deceleration of the shower front, and study its properties.
\end{abstract}


\maketitle

\section{Introduction}

Energy cosmic rays with energies $> 10^{15}$ eV cannot be directly measured due to their
low flux, so they are detected via the particle showers they create when they enter
the atmosphere. The secondary particles present in the extensive air shower (EAS) can
give us information on the primary cosmic ray, such as its arrival direction, its energy and,
more recently, its composition \cite{yushkov,report}. There are three main detection
techniques for inspecting the EAS and elucidating the properties of the primary particle. 
First, there is the surface detection technique,
wherein the shower particles arriving at ground level can be
detected, and this sample of the shower is then used to analyse the properties of the primary.
Second, there is the fluorescence technique. Whenever a particle shower passes through
the atmosphere, the nitrogen molecules within it become excited and emit fluorescence light
when they de-excite. This fluorescent light can be detected in moonless nights, and its
luminosity is proportional to the number of charged particles traversing a given region of
the atmosphere, allowing to reconstruct the development of the EAS. And third, we have
the radio technique.

The radio detection technique of high-energy particles consists in detecting, by
the use of radio antennas, the electric field created by the charged particles present
in a particle shower. The amplitude, ground footprint, arrival time and polarization of the electric
field can be correlated to the arrival direction, energy and even composition of
the primary particle \cite{lofarnature,flothesis}.
The radio technique is a well-established technique as of today \cite{frankreview}.
This technique was proposed in 1962 by G. Askaryan as a method for detecting cosmic rays
and neutrinos \cite{askaryan}. It did not take long to observe pulses correlated with
showers measured by
air shower arrays \cite{allan}, but it was abandoned due to technical limitations
that prevented it from being as competitive as the surface detection. With the advent of
modern electronics, radio detection and the interest it generated were reawakened.

We know today that the electric field created by EAS can be explained by the superposition
of two effects. The dominant effect is called the \emph{geomagnetic effect} and it is induced
by the deflection of the charged particles by the geomagnetic field. Positive and negative
charges deviate in opposite directions, creating a total current in the same direction that
produces an electromagnetic field with a 
{polarization along the Lorentz force direction,
$\mathbf{v}\times\mathbf{B}$, with $\mathbf{v}$ the shower axis and 
$\mathbf{B}$ the geomagnetic field.
The \emph{Askaryan effect} is related to the apparition of an excess of negative charge
(electrons) within the shower, which creates detectable electric field with a polarization
radial with respect to the shower axis. We can find in
\cite{washington} a theoretical description of these two effects and in 
\cite{codaxs,radioemissionprd}
data that evidence their existence and interplay to create the total electric field.
Since a significant part of the shower particles travel faster than the speed of light,
relativistic interference effects are also relevant, leading to the apparition of a Cherenkov
cone.

Most radio experiments detecting air showers nowadays (for instance, 
CODALEMA \cite{icrcarxiv}, AERA \cite{aera}, Tunka-Rex \cite{tunka} or LOFAR
\cite{lofar}) measure the electric field in the $[20-80]$ MHz band. 
This is due to the fact that the field from EAS is usually coherent up to several
tens of MHz, and above $80$ MHz we find FM radio emitters. However, near
the Cherenkov angle, the field can be coherent up to GHz frequencies
\cite{reflected}. There are some complications for measuring below $20$ MHz,
mainly the atmospheric and galactic noise, and that is why these experiments
have chosen the aforementioned band. 

Several experiments in the past have informed of the observation of pulses
correlated with air showers at frequencies below $20$ MHz, but the results are not
well understood. The first detection of this kind took place as early as in 1967
\cite{r7}. In 1970, researches at Haverah Park found an average low-frequency
signal (at $3.6$ MHz) of $300 \ \mu$V m${}^{-1}$ MHz${}^{-1}$, 
averaged over 400 showers \cite{allanlow}. However, a repeat run gave a
null result \cite{r10}. In 1971 \cite{r10}, a signal of about
$1\ \mu$V m${}^{-1}$ MHz${}^{-1}$ at $1.98$ MHz was measured.
According to their estimation, the shower energy was $\sim 2\times 10^{14}$ eV,
which is incredibly low for a radio-detected shower. This energy estimation
allows them to conclude that a $10^{17}$ eV shower should present a
field strength of $500 \ \mu$V m${}^{-1}$ MHz${}^{-1}$, which seems
implausible given the energy measurement.
In \cite{r8}, a mean field of $2.5\pm 0.6 \ \mu$V m${}^{-1}$ MHz${}^{-1}$
was measured at $3.6$ MHz, and with the help of the shower size estimation
and the hypothesis of pure geomagnetic emission, they extrapolate
$300 \ \mu$V m${}^{-1}$ MHz${}^{-1}$ at $10^{17}$, more than one order of
magnitude larger than the standard $20$-$80$ MHz emission. We must
point that the measured fields so far had been of the order of a few 
$\mu$V m${}^{-1}$ MHz${}^{-1}$. In fact, \cite{allanlow2} reports an
upper limit of $0.6 \ \mu$V m${}^{-1}$ MHz${}^{-1}$ for a $10^{17}$ eV
shower, obtained with an antenna located $\lambda/10$ wavelengths above ground,
which is incompatible with previous estimations.

In spite of the controversy, later experiments reported direct measurements
of large electric fields.
The Akeno experiment \cite{r18} pursued even lower frequencies ($26$-$300$ kHz)
and found correlation of signals in that band with signals at higher frequencies, with
a typical amplitude of hundreds of $\mu$V m${}^{-1}$ MHz${}^{-1}$.
A measure they repeated in \cite{r22}.
EAS-RADIO \cite{r23}, set up at EASTOP was able to measure again fields
of the same order.

A discussion of these experiments along with some more and a brief theoretical
summary can be found in \cite{sdpicrc}. The experiments and analyses cited
therein agree that the standard calculations used in radio experiments are not
enough to describe the measured field. Some mechanisms, such as transition
radiation or deviation of particles in the geoelectric field, do not seem to be
favored by the data \cite{r18,r24}. Coherent deceleration of the shower front,
also called sudden death pulse (SDP) has been proposed in \cite{r18,sdpicrc,sdparxiv}
as a means to explain the low-frequency emission.

In order to measure again the electric field of an EAS at low frequencies and shed
light on the emission mechanisms, the EXTASIS experiment has been set up at the
Nan\c{c}ay radio observatory. With a frequency band spanning from $1.7$ to $3.7$
MHz and seven antennas with vertical and horizontal polarizations hung at the top of
$9$ m high masts, EXTASIS has already detected low-frequency pulses correlated
with pulses in the $[20-80]$ MHz band and with events seen by a 
surface detector \cite{antony}. The data look like a real cosmic ray event, and
the reconstruction of the arrival direction 
in the $[1.7-3.7]$ MHz band is in agreement with the reconstruction in the
$[20-80]$ MHz band and the one given by the surface detector.
These data also hint at a
larger reach of the electric field at low frequencies. EXTASIS has also among its goals
the detection of the sudden death pulse. \\
One of the problems that arise is the correct calculation of the electric field.
Most of the theoretical formulas \cite{zhs,buniy} and standard codes used, such as
SELFAS \cite{selfas}, ZHAireS \cite{zhaires} or CoREAS \cite{coreas}, use the
far-field approximation to obtain the radiation field only. The far-field approximation
holds when the wavenumber $k$ and the distance from emitter to observer
verifies $kR = 2\pi R/\lambda \gg 1$, which allows us to approximate the electric
field as a pure radiation field. At $\sim 1$ MHz, the wavelength is $\sim 300$ m,
which is comparable to the distance from the measuring antenna to the shower
core and axis. If, moreover, we expect a noticeable emission from the shower
close to the core (as in the sudden death mechanism), the far-field approximation
could not suffice. In particular, for an antenna at $100$ m from the shower core,
$kR = 2\pi\cdot 100/300 \sim 2$, so near-field effects could be important
at the EXTASIS frequencies. \\
We will begin by deriving in Section~\ref{sec:time}
a time-domain formula for the field of a particle track valid for all
frequencies. 
%and that therefore allows us to go beyond the far-field approximation.
We will derive again the formula in frequency domain in Section~\ref{sec:freq}.
The time-domain and frequency-domain formulas will be compared to
the far-field approximation or ZHS formula \cite{zhstime}. 
In Section~\ref{sec:mc} we will explain the implementation of the
exact formula in the SELFAS Monte Carlo code.
In Section~\ref{sec:sim} we will
show some examples of the predictions of this formula for the emission coming from
EAS and give a range of validity for the far-field approximation. We will introduce
the coherent deceleration of the shower front or sudden death pulse and study
its properties in Section~\ref{sec:sdp}.

\section{Electric field of a particle track in time domain}
\label{sec:time}
\subsection{Derivation of the formula in time domain}

Our aim is to derive a formula for the electric field emitted by a particle contained
inside a shower. Since we are operating at the quantum scale, we must justify the
use of classical electrodynamics. The trajectory of the particle inside a particle
shower is created by the interactions with the medium that cause the particle to change
direction and speed almost instantaneously. Physically, however,
there must be a
time related to the interaction below which the use of quantum
theory is imperative. Let us call this time $\Delta t_q$. We expect the frequency
components of the field to be valid until the observation frequency is comparable
to $1/\Delta t_q$, and in this low-frequency region we can apply classical 
electrodynamics \cite{peskin}.
Let us assume $\Delta t_q$ is related to the Compton wavelength $\lambda_C$ of
the electron,
\begin{equation}
\Delta t_q \sim \frac{h}{m_e c^2} = \frac{\lambda_C}{c}
= \frac{2.425\cdot 10^{-12} \ \textrm{m}}{0.3 \ \textrm{m} \ \textrm{ns}^{-1}} =
8.09\cdot 10^{-12} \ \textrm{ns},
\end{equation}
which implies that the order of magnitude of the maximum frequency where the
classical approach should work is
\begin{equation}
\nu_\textrm{classical} \ll \frac{1}{\Delta t_q} \sim
\frac{m_e c^2}{h} = \frac{c}{\lambda_C} \sim 1.24\cdot 10^{11} \ \textrm{GHz}.
\end{equation}
We are interested in radio emission below the THz, so the use of classical electrodynamics
is justified.

We begin with Maxwell's equations for the potentials inside an homogeneous, dielectric medium
with permittivity $\epsilon$, independent of the frequency. 
Its refractive index is, therefore, $n = \sqrt{\epsilon/\epsilon_0}$,
and the speed of light in the medium is $c_n = c/n$. Using the Lorenz gauge,
\begin{eqnarray}
\label{eq:phi}
\nabla^2 \Phi - \frac{1}{c_n^2}\frac{\partial^2 \phi}{\partial t^2} & = & - \frac{\rho}{\epsilon} \\
\nabla^2 \mathbf{A} - \frac{1}{c_n^2}\frac{\partial^2 \mathbf{A}}{\partial t^2} & = & - \mu_0 \mathbf{J},
\label{eq:A}
\end{eqnarray}
where $t$ is the observer time, $\Phi$ and $\mathbf{A}$ are the electromagnetic potentials,
$\rho$ is the charge density and $\mathbf{J}$ is the current density. Eqs.~\eqref{eq:phi}
and~\eqref{eq:A} are similar to their vacuum counterparts. Formally, we have made the changes
$\epsilon_0 \rightarrow \epsilon$ and $c \rightarrow c_n$, so the formal solutions for the
potential are the usual retarded potentials with these substitutions:
\begin{eqnarray}
\label{eq:phi3}
\Phi(\mathbf{x}, t) = \frac{1}{4\pi\epsilon} \int \textrm{d}^3 x' \frac{1}{R}
[\rho(\mathbf{x'}, t')]_\textrm{ret} \\
\mathbf{A}(\mathbf{x}, t) = \frac{\mu_0}{4\pi} \int \textrm{d}^3 x' \frac{1}{R}
[\mathbf{J}(\mathbf{x'}, t')]_\textrm{ret},
\label{eq:A2}
\end{eqnarray}
where $R = |\mathbf{x} - \mathbf{x}'|$ and $t'$ is the \emph{retarded} or \emph{emission}
time, defined implicitely as
\begin{equation}
t' + \frac{1}{c_n} |\mathbf{x} - \mathbf{x}'(t') | = t
\end{equation}
As long as our source
is not superluminous, that is, as long as its speed is less than $c_n$, the speed of light in
the medium, the calculation of the field is as straighforward as in the vacuum case. In the
case of superluminous motion, we can still use Eqs.~\eqref{eq:phi3} and~\eqref{eq:A2} to
derive the fields, being aware that we can have several retarded times for a single observation
time: $t'(t)$ is a multivalued function. For the superluminous case, Jefimenko's
equation for the electric field is still valid, and can be written as
\begin{equation}
\mathbf{E}(\mathbf{x}, t) = \frac{1}{4\pi\epsilon} \int \textrm{d}^3 x'
\left[
\frac{\hatR}{R^2}[\rhoprim]_\textrm{ret} + 
\frac{\hatR}{c_n R}\left[ \frac{\partial\rhoprim}{\partial t'} \right]_\textrm{ret} -
\frac{1}{c_n^2 R} \left[ \frac{\partial\Jprim}{\partial t'} \right]_\textrm{ret}
\right],
\label{eq:jef1}
\end{equation}
where $\hatR = \frac{\xdiff}{|\xdiff|}$. Since the integrands are functions of $\mathbf{x}$,
$\mathbf{x'}$, and $t$ we can cast the following property:
\begin{equation}
\left[
\frac{\partial f \xtprim}{\partial t'}
\right]\ret = 
\frac{\partial}{\partial t} [f\xtprim]\ret \frac{\partial t(t',\mathbf{x},\mathbf{x'})}{\partial t'} =
\frac{\partial}{\partial t} [f\xtprim]\ret,
\end{equation}
which, along with the fact that $R = R(\mathbf{x},\mathbf{x'})$ inside the integral 
(it is not a fuction of $t$ or $t'$),
allows us to rewrite Eq.~\eqref{eq:jef1} in the following way:
\begin{equation}
\mathbf{E}(\mathbf{x}, t) = \frac{1}{4\pi\epsilon} \left [ \int \textrm{d}^3 x'
\frac{\hatR}{R^2}\left[ \rhoprim \right]_\textrm{ret} + 
\frac{\partial}{\partial t} \int \textrm{d}^3 x'
\frac{\hatR}{c_n R}\left[ \rhoprim \right]\ret -
\frac{\partial}{\partial t} \int \textrm{d}^3 x'
\frac{1}{c_n^2 R} \left[ \Jprim \right]\ret
\right].
\label{eq:jef2}
\end{equation}
We will proceed now with the definition of our source. We will consider a particle track, which
is the building block for all Monte Carlo codes that calculate the electric field of an EAS within
the microscopic approach. 
Let us consider a neutral atom or molecule in which all the particles are located
approximately in the same point so that the charge density in all of space is zero 
before the time $t_1$.
At $t_1$, a point charge with charge $+q$ separates from the particle with charge
$-q$ at the position $\mathbf{x}_1$. The $-q$ charge
stays at the position $\mathbf{x_1}$, while the charge $+q$ travels
in a straight line at constant speed until it is suddenly stopped at $t_2$, at the position
$\mathbf{x}_2$. This charge density can be written as
\begin{eqnarray}
\rho\xtprim = & - & q \deltaprim{1} \thetaprim{1}  \nonumber \\
                  & + &  q\deltav [\rect] \nonumber \\
                  & + & q \deltaprim{2} \thetaprim{2},
                  \label{eq:charge}
\end{eqnarray}
where we have used the Dirac delta function $\delta$ and the Heaviside step function $\Theta$.
The current density is, on the other hand,
\begin{equation}
\Jprim = q \mathbf{v} \deltav [\rect].
\label{eq:current}
\end{equation}
Eq.~\eqref{eq:charge} guarantees that charge is conserved. An electric field calculated with a
non-conserved charge is not a solution of Maxwell's equations, since their definition implies
the conservation of charge. Not taking into account the first and third terms in the RHS of
Eq.~\eqref{eq:charge} results in an unphysical radiation field, for instance (as shown in 
Section~\ref{sec:timecomp}). 
Note that we have chosen a $-q$ charge in order to balance out the total
charge contained in the space, but we could have used a single $+q$ charge for our problem,
by substituting the first term in Eq.~\eqref{eq:charge} by $+q \deltaprim{1} \Theta(t_1 - t')$,
and the resulting electric field would only differ in a static Coulomb term.

The integrals containing a $\deltaprim{1}$ or $\deltaprim{2}$ are integrated trivially. However,
the term $\deltav$ must be written as
\begin{eqnarray}
\delta^3(\mathbf{g}(\mathbf{x}')) & = & \delta^3 \left(
\mathbf{x}' - \mathbf{x}_1 - \mathbf{v} \left (
t - \frac{|\xdiff|}{c_n} - t_1
\right ) \right) = 
\sum_i \frac{\delta^3 (\mathbf{x}' - \mathbf{x}_{p,i} \xt)}{\left| \frac{\partial \mathbf{g}}{ \partial \mathbf{x}' } \right|_{\mathbf{x}_{p,i}}}
\nonumber \\
& = & \sum_i \frac{\delta^3 (\mathbf{x}' - \mathbf{x}_{p,i} \xt)}{\left| 1 - \mathbf{v}\cdot\hatR(\mathbf{x}_{p,i} \xt)/c_n \right|}
= \sum_i \frac{\delta^3 (\mathbf{x}' - \mathbf{x}_{p,i} \xt)}{\kappa_i},
\label{eq:deltak}
\end{eqnarray}
where $\mathbf{x}_{p,i} \xt$ is the retarded position of the particle as a function of $\xt$. These
deltas are going to ensure that $\mathbf{x}'$ is only evaluated at the retarded position, so upon
integration we must substitute $\mathbf{x}'$ for the retarded particle position and introduce
the $\kappa$ factors, where we have defined
\begin{equation}
\kappa_i = \left| 1 - \mathbf{v}\cdot\hatR(\mathbf{x}_{p,i} \xt)/c_n \right| =
 \left| 1 - \mathbf{v}\cdot\hatR_i/c_n \right|\ret
\end{equation}
The index $i$ runs through the several possible retarded positions, that is, the
roots of $\mathbf{g}(\mathbf{x}') = 0$. $\kappa$ is, in fact, related to the derivative
of the observation time $t$ with respect to the retarded time $t'$:
\begin{equation}
\frac{\mathrm{d}t}{\mathrm{d}t'} = 1 - \mathbf{v}\cdot\hatR/c_n
\end{equation}
Let $\theta$ be the angle formed by $\mathbf{v}$ and $\hatR$, that is, the angle that
the particle-observer line makes with the track.
Let us define the Cherenkov angle $\theta_C$ as the angle that makes this derivative zero.
\begin{equation}
1 - \mathbf{v}\cdot\hatR/c_n = 1 - n\beta\cos\theta_C = 0,
\end{equation}
that is,
\begin{equation}
\theta_C = \mathrm{acos} \left( \frac{1}{n\beta} \right),
\end{equation}
where $\beta = v/c$. If, for our case, the track is always seen with an angle $\theta > \theta_C$,
$\frac{\mathrm{d}t}{\mathrm{d}t'} > 0$, so $t$ is monotonically increasing and there is only
one $t'$ for a given $t$. The same is true if $\theta < \theta_C$, although in this case
the derivative is $\frac{\mathrm{d}t}{\mathrm{d}t'} < 0$. However, if for a point along the track
$\theta = \theta_C$ is verified, 
we can have two retarded times for a single observation time. Knowing
this, we can drop the sum over $i$ and introduce the retardation brackets, keeping in mind
that we may have two retarded positions below and above the Cherenkov angle 
whose fields we have to add.

Using Eq.~\eqref{eq:deltak} in Eq.~\eqref{eq:jef2} with the sources specified in
Eqs.~\eqref{eq:charge} and~\eqref{eq:current}, the field can be written as
\begin{eqnarray}
\mathbf{E}\xt & = & \frac{q}{4\pi\epsilon} \Bigg\{
- \frac{\hatR_1}{R_1^2} [\thetaprim{1}]\ret - \frac{\hatR_1}{c_n R_1}\frac{\partial}{\partial t} [\thetaprim{1}]\ret
\nonumber \\
& + &  \left[ \frac{\hatR}{\kappa R^2} (\Rect) \right]\ret
+ \frac{1}{c_n}\frac{\partial}{\partial t} \left[ \frac{\hatR}{\kappa R} (\Rect) \right]\ret
-  \frac{\mathbf{v}}{c_n^2}\frac{\partial}{\partial t} \left[ \frac{1}{\kappa R} (\Rect) \right]\ret
\nonumber \\
& + & \frac{\hatR_2}{R_2^2} [\thetaprim{2}]\ret + \frac{\hatR_2}{c_n R_2}\frac{\partial}{\partial t} [\thetaprim{2}]\ret
\Bigg\},
\label{eq:Etime1}
\end{eqnarray}
having defined the boxcar function
\begin{equation}
\Rect = \rect
\end{equation}
and $R_{1,2}$, $\hatR_{1,2}$ being the distances and unit vectors from the first and last point
of the track to the observer. 
In order to simplify the expression, we can derivate the terms $\frac{\partial}{\partial t} [\thetaprim{i}]\ret$.
The step function is always $0$ or $1$, either using retarded time or observation time, so its derivative
with respect to observation time must be an unit Dirac delta, but at the 
observation time $t$ corresponding to the
retarded time $t' = t_i$:
\begin{equation}
\frac{\partial}{\partial t} [\thetaprim{i}]\ret = 
\left[ \frac{\partial}{\partial t'} \thetaprim{i} \right]\ret =
[\delta(t'-t_i)]\ret,
\end{equation}
which allows us to rewrite Eq.~\eqref{eq:Etime1} as
\begin{eqnarray}
\mathbf{E}\xt & = & \frac{q}{4\pi\epsilon} \Bigg\{
- \frac{\hatR_1}{R_1^2} \Theta(t - t_1 - R_1/c_n) - \frac{\hatR_1}{c_n R_1} \delta(t - t_1 - R_1/c_n)
\nonumber \\
& + &  \left[ \frac{\hatR}{\kappa R^2} (\Rect) \right]\ret
+ \frac{1}{c_n}\frac{\partial}{\partial t} \left[ \frac{\hatR}{\kappa R} (\Rect) \right]\ret
-  \frac{\mathbf{v}}{c_n^2}\frac{\partial}{\partial t} \left[ \frac{1}{\kappa R} (\Rect) \right]\ret
\nonumber \\
& + & \frac{\hatR_2}{R_2^2} \Theta(t - t_2 - R_2/c_n) + \frac{\hatR_2}{c_n R_2} \delta(t - t_2 - R_2/c_n)
\Bigg\}.
\label{eq:Etime2}
\end{eqnarray}
We have used the fact that $t' = t - R_i/c_n$ before $t' = t_1$ and after $t' = t_2$.
The first line in Eq.~\eqref{eq:Etime2} contains a static Coulomb field turned on at $t_1$ and an
impulse radiation field. The third line shows a similar field, but corresponding to the instant when
the particle stops. The second line is analogous to the Heaviside-Feynman expression of the electric
field of a particle existing for a limited time, used in \cite{selfas} for the calculation of the electric field.
The impulse fields with the form
\begin{equation}
\frac{\hatR_i}{c_n R_i}\delta(t-t_i-R_i/c_n)
\end{equation}
come directly from the imposition of charge conservation, and they represent a radiation field that
would be missing had we not taken into account a realistic charge density. This field
is created by the changes in the charge density, and so we can pair it with the second
term in the second line of of Eq.~\eqref{eq:Etime2}, since they must always appear together.
Let us develop what happens at the time
$t_{1,\mathrm{obs}} = t_1 + R_1/c_n$ with these two terms:
\begin{equation}
\mathbf{E}_\textrm{cons} (\mathbf{x}, t) =
\frac{q}{4\pi\epsilon c_n}\left[
-\frac{\hatR_1}{R_1} +
\frac{\hatR_1}{R_1\left| 1-\mathbf{v}\cdot\hatR_1/c_n \right|}  \right]
\delta(t-t_1-R_1/c_n).
\label{eq:Econs}
\end{equation}
The second term on the RHS in Eq.~\eqref{eq:Econs} cannot exist on its own. It represents a sudden apparition
of charge, which is not possible without a creation of an opposite charge at the same point, 
which in turns creates an electric field given by the first term.
Analogously for the deceleration of our particle, we need to have a field of the kind:
\begin{equation}
\mathbf{E}_\textrm{cons} (\mathbf{x}, t) = \frac{q}{4\pi\epsilon c_n} \left[
\frac{\hatR_2}{R_2} -
\frac{\hatR_2}{R_2\left| 1-\mathbf{v}\cdot\hatR_2/c_n \right|} \right]
\delta(t-t_2-R_2/c_n),
\label{eq:Econs2}
\end{equation}
where the first term on the RHS is reminding us that the particle does not disappear but instead stops
and stays at the same place.

\subsection{Far-field approximation. The ZHS formula}

Eq.~\eqref{eq:Etime2} can be taken to the far-field limit arriving at the time-domain
ZHS formula \cite{zhstime}. Let us begin by dropping all the terms with a $1/R^2$ dependence
in Eq.~\eqref{eq:Etime2}.
\begin{eqnarray}
\mathbf{E}\xt & = & \frac{q}{4\pi\epsilon} \Bigg\{
 - \frac{\hatR_1}{c_n R_1} \delta(t - t_1 - R_1/c_n) + \frac{\hatR_2}{c_n R_2} \delta(t - t_2 - R_2/c_n)
 \nonumber \\
& + & \frac{1}{c_n R}\frac{\partial}{\partial t} \left[ \frac{\hatR}{\kappa} (\Rect) \right]\ret
-  \frac{\mathbf{v}}{c_n^2 R}\frac{\partial}{\partial t} \left[ \frac{1}{\kappa} (\Rect) \right]\ret
\Bigg\}.
\end{eqnarray}
Let us suppose now that the unit vector $\hat{R}$ and $\kappa = |1-n\beta\cos\theta|$ do not change
during the track flight, which is equivalent to have the observation angle $\theta$ limited to a narrow range
or having the track far away from the observers. Let us take $R \approx R_0$, with $R_0$ one of
the points in the track, for instance, the center point.
Partials time derivatives are then zero except when
the particle accelerates and decelerates:
\begin{eqnarray}
\mathbf{E}\xt & = & \frac{q}{4\pi\epsilon} \Bigg\{
 - \frac{\hatR_0}{c_n R_0} \delta(t - t_1 - R_1/c_n) + \frac{\hatR_0}{c_n R_0} \delta(t - t_2 - R_2/c_n)
 \nonumber \\
& \pm & \frac{1}{c_n }\frac{\hatR_0}{\kappa R_0} \delta(t - t_1 - R_1/c_n)
\mp  \frac{\mathbf{v}}{c_n^2 \kappa R_0} \delta(t - t_1 - R_1/c_n)
\nonumber \\
& \mp & \frac{1}{c_n }\frac{\hatR_0}{\kappa R_0} \delta(t - t_2 - R_2/c_n)
\pm  \frac{\mathbf{v}}{c_n^2 \kappa R_0} \delta(t - t_2 - R_2/c_n)
\Bigg\},
\label{eq:prezhs1}
\end{eqnarray}
where from the $\pm$ and $\mp$ symbols, we take the upper one if $\theta > \theta_C$ and
the lower one if otherwise. Since we are in the far field, we can use the Fraunhofer approximation
for the distance: choosing a point $\mathbf{x}_0$ inside the track, and supposing the particle
is at $\mathbf{x}_0$ when $t' = t_0 = 0$,
\begin{equation}
R_i = | \mathbf{x} - \mathbf{x}_0 + \mathbf{v}t' |
\approx R_0 - \mathbf{v}\cdot\hatR_0 t_i =
R_0 - \cos\theta \ v t_i.
\label{eq:fraunhofer}
\end{equation}
Plugging Eq.~\eqref{eq:fraunhofer} into Eq.~\eqref{eq:prezhs1} and renaming $R_0 \rightarrow R$,
\begin{eqnarray}
\mathbf{E}\xt & = & \frac{q}{4\pi\epsilon} \Bigg\{
 - \frac{\hatR}{c_n R} \delta \left(t - \frac{nR}{c} - (1-n\beta\cos\theta) t_1\right) 
 + \frac{\hatR}{c_n R} \delta \left(t - \frac{nR}{c} - (1-n\beta\cos\theta) t_2 \right)
 \nonumber \\
& \pm & \frac{1}{c_n }\frac{\hatR}{\kappa R} \delta \left(t - \frac{nR}{c} - (1-n\beta\cos\theta) t_1\right) 
\mp  \frac{\mathbf{v}}{c_n^2 \kappa R} \delta \left(t - \frac{nR}{c} - (1-n\beta\cos\theta) t_1\right) 
\ \\
& \mp & \frac{1}{c_n }\frac{\hatR}{\kappa R} \delta \left(t - \frac{nR}{c} - (1-n\beta\cos\theta) t_2 \right) 
\pm  \frac{\mathbf{v}}{c_n^2 \kappa R} \delta \left((t - \frac{nR}{c} - (1-n\beta\cos\theta) t_2) \right)
\Bigg\}. \nonumber
\label{eq:prezhs2}
\end{eqnarray}
We now make use of the following identity:
\begin{eqnarray}
\hatR - \frac{\hat R}{\kappa} + \frac{\mathbf{v}}{\kappa c_n} & = &
\pm \frac{1}{\kappa} \left[ \frac{\mathbf{v}}{c_n} -n\beta\hatR\cos\theta \right] =
\pm \frac{1}{\kappa c_n} \left[ \mathbf{v} (\hatR\cdot\hatR)
- \hatR(\mathbf{v}\cdot\hatR) \right] \nonumber \\
& = & \pm \frac{1}{\kappa c_n} \left[ -\hatR\times(\hatR\times\mathbf{v}) \right]
= \pm \frac{\mathbf{v_\perp}}{\kappa c_n},
\label{eq:vperp}
\end{eqnarray}
where we take the plus sign if $\theta > \theta_C$ and the minus sign if otherwise.
$\mathbf{v_\perp}$ is, by definition, the projection of the particle velocity perpendicular to the
line joining the particle and the observer, also called the \emph{line of sight}. With
Eq.~\eqref{eq:vperp} and Eq.~\eqref{eq:prezhs2}, using a positron charge
defined as $q = e > 0$ and using 
$1/(\epsilon c_n^2) = \mu_r/(\epsilon_0 c^2)$, we arrive at the same equation
than in \cite{zhstime}:
\begin{equation}
\mathbf{E}_\textrm{ZHS} \xt =
-\frac{e\mu_r}{4\pi\epsilon_0 c^2 R} \mathbf{v}_\perp
\frac{\delta \left(t - \frac{nR}{c} - (1-n\beta\cos\theta) t_1\right)  - \delta \left(t - \frac{nR}{c} - (1-n\beta\cos\theta) t_2 \right)}{1-n\beta\cos\theta}.
\label{eq:zhs}
\end{equation}
Note that the $\pm$ and the $\mp$ have been substituted by the 
sign of the denominator that
is positive outside the Cherenkov cone and negative inside.

\subsection{Calculation of time-domain field and comparison with far-field (ZHS) formula}
\label{sec:timecomp}

\begin{figure}
\includegraphics[width=0.5\textwidth]{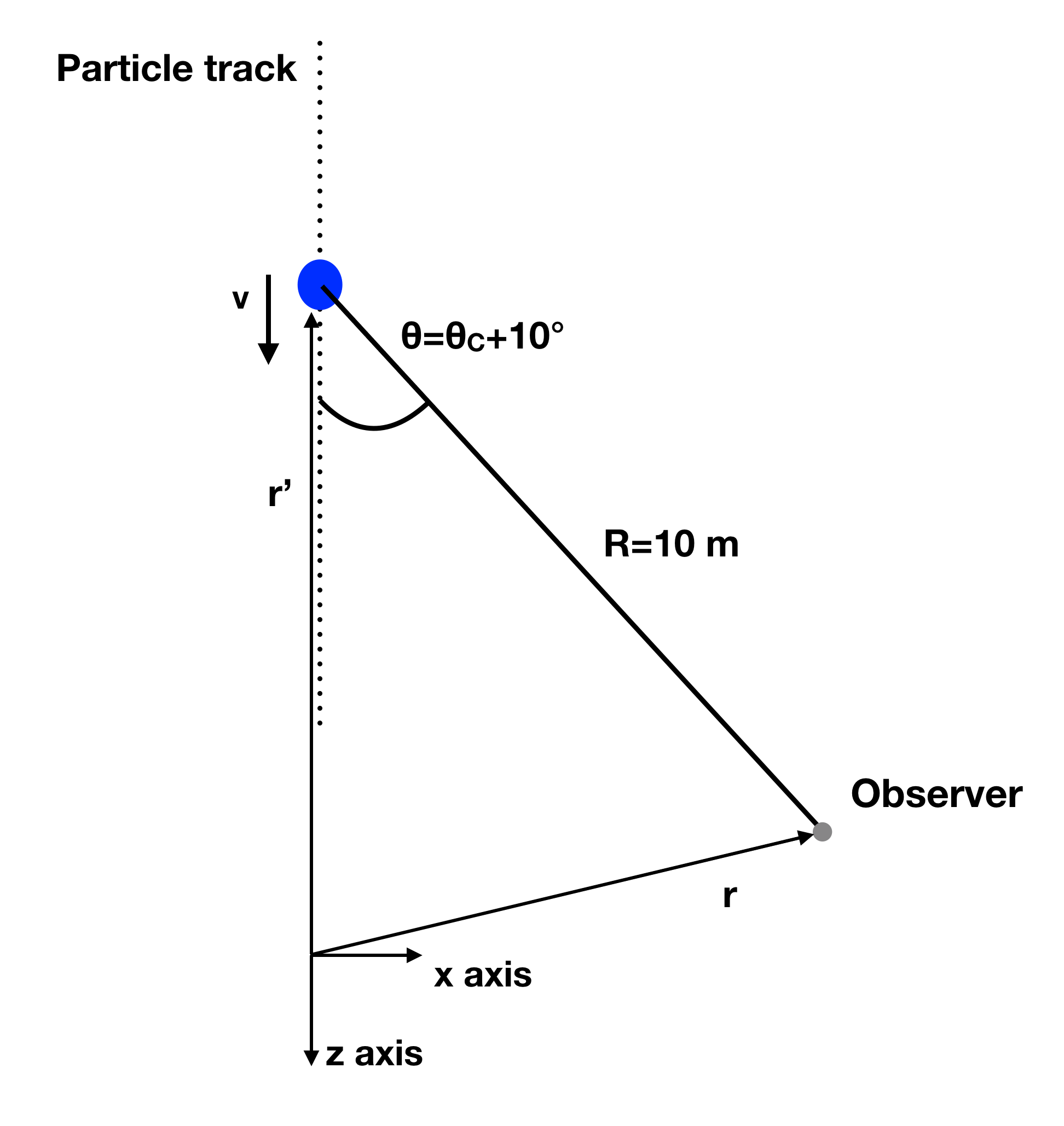}
\caption{Sketch containing the particle track and the observer. $\theta$ is the angle
formed by the particle velocity $\mathbf{v}$ and the line joining the particle and the observer.
The observer lies at $\mathbf{r}$ and the particle at $\mathbf{r'}$. $R$ is the distance
between observer and particle. The particle travels along the $z$ axis (indicated in
the figure along with the $x$ axis).}
\label{fig:schema}
\end{figure}

Let us consider an electron track ($q = -e$) along with a positive charge ($q = e$) that stays at rest.
The time-domain electric field of such a configuration can be calculated numerically using Eq.~\eqref{eq:Etime2}.
Even though one of the main aims of this work is to develop an equation to calculate the field created
by an extensive air shower, Eq.~\eqref{eq:Etime2} is valid for dense dielectric media as well. Due to this, and
also to the fact that superluminous effects (like the Cherenkov cone) are more prominent in dense media, we
will choose a dielectric medium with a refractive index of $n = 1.78$, akin to that of deep Antarctic ice, to
perform our calculations and to better illustrate our point. 
Let us embed a $1.2$ m long electron track traveling along
the $z$ axis with a speed of $v\sim c$ 
in the aforementioned medium and place an observer at $10$ m and at an angle of 
$\theta_C + 10^\circ = 65.82^\circ$, with $\theta_C$ the Cherenkov angle
(see Fig.~\ref{fig:schema}). 
In order to evaluate Eq.~\eqref{eq:Econs2}, we choose a numerical step 
$\Delta t = 0.05$ ns and
define our numerical derivative at a point $t_i$ as a two-point derivative as follows:
%\{red}{What the fuck is going on with this spacing?}
\begin{equation}
\frac{\mathrm{d}f}{\mathrm{d}{t}}(t_0) \approx \frac{f(t_0 + \Delta t) - f(t_0 - \Delta t)}{2\Delta t}
\end{equation}
Besides evaluating it, we can compare our formula with the ZHS algorithm, which has been
proven \cite{prd87} to give the correct results for the far-field regime, 
that is, $kR \gg 1$, with $k$ the
wavenumber and $R$ the distance between particle and observer.
We show the results in Fig.~\ref{fig:tracktime}. We find that the ZHS algorithm (blue points) and
Eq.~\eqref{eq:Etime2} (red line) give the same result, which is understandable knowing that the
field lasts for several nanoseconds, which means that 
for the frequencies contributing the most to the radiation field,
$kR \sim 2\pi \frac{R}{c_n T} \sim 370$, with
$T$ being the inverse of the frequency $\nu$, of the order of the GHz. However, if we do not
take into account the terms in Eqs.~\eqref{eq:Econs} and~\eqref{eq:Econs2} that guarantee the
conservation of charge, we are not able to reproduce the impulses corresponding to the acceleration
and deceleration of the particle (green lines), meaning that such terms cannot
be omitted for a physical
calculation of the electric field. Putting an observer at $100$ m from the track, where the
acceleration and deceleration fields are more important than in the previous case, demonstrates that not including
this terms results in an incorrect impulse field, see Fig.~\ref{fig:tracktime2}.

\begin{figure}
\includegraphics[width=0.75\textwidth]{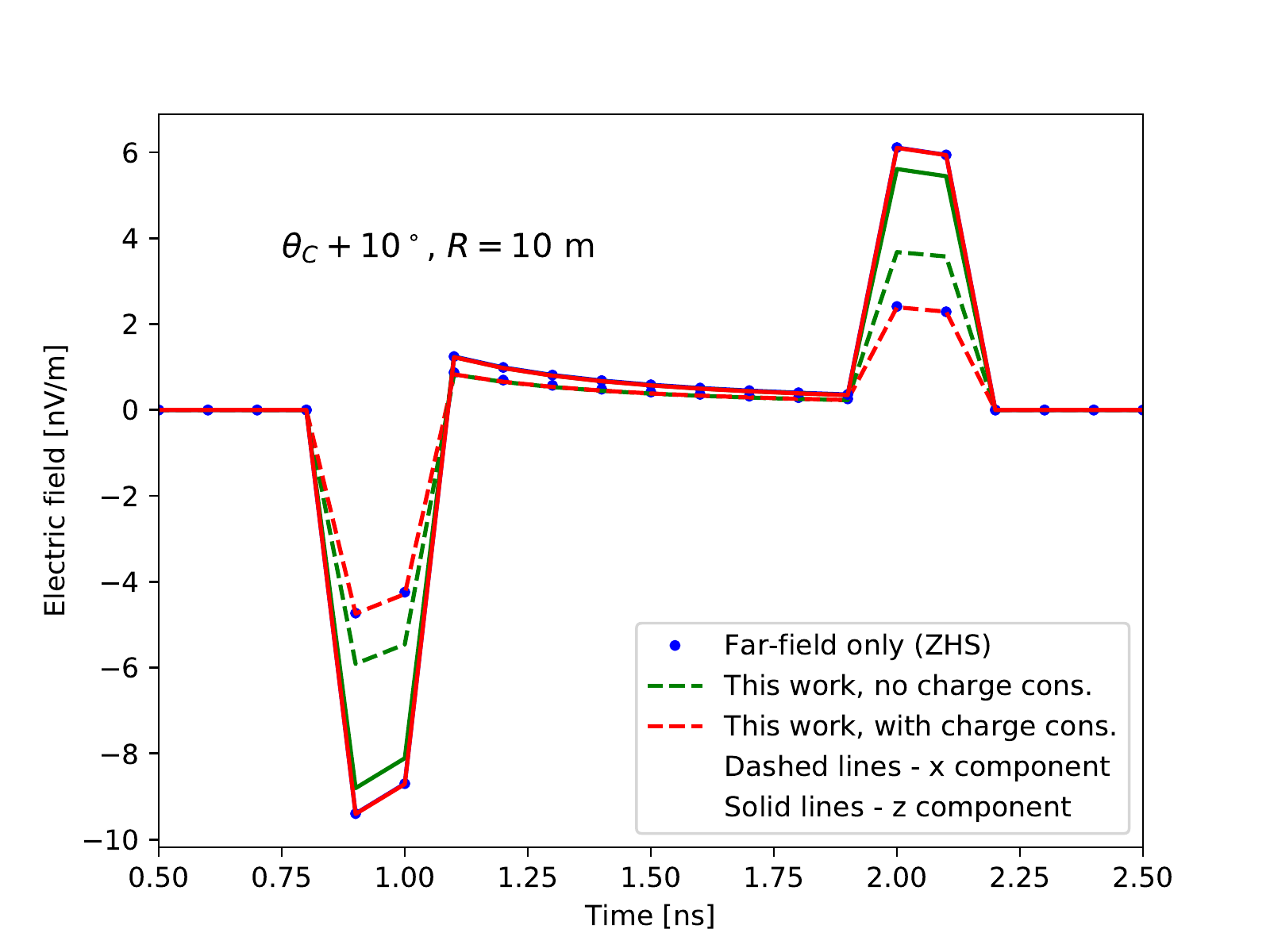}
\caption{Electric field as a function of time created by a $1.2$ m long electron track traveling along
the $z$ axis in deep Antarctic ice ($n = 1.78$).
The observer is placed at $\theta_C + 10^\circ$ and at a distance of $R = 100$ m. Red lines represent
Eq.~\eqref{eq:Etime2}, blue points represent the ZHS algorithm and green lines represent Eq.~\eqref{eq:Etime2}
but without the terms needed for the conservation of charge. The numerical time step is $\Delta t = 0.1$ ns.}
\label{fig:tracktime}
\end{figure}

\begin{figure}
\includegraphics[width=0.75\textwidth]{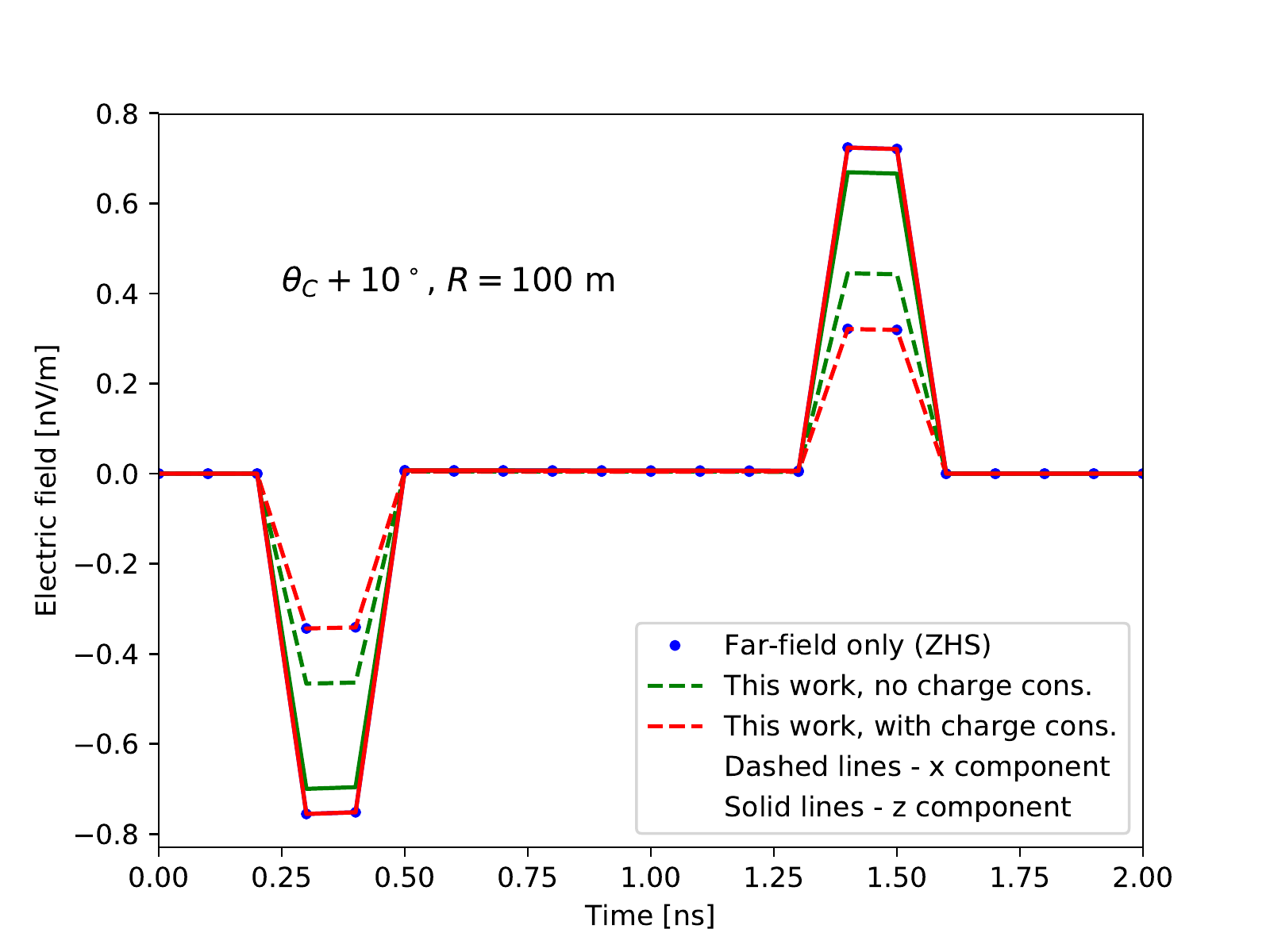}
\caption{Same as Fig.~\ref{fig:tracktime} but with an observer at $R = 100$ m.}
\label{fig:tracktime2}
\end{figure}

We show in Fig.~\ref{fig:tracktime3} that our formula is also valid for calculating the field inside the Cherenkov
cone. The predicted polarity is in agreement with the polarity from the ZHS formula. Besides, outside (Fig.~\ref{fig:tracktime}) and inside
the Cherenkov (Fig.~\ref{fig:tracktime3}) cone we find the same polarity. Let us remember that the acceleration and deceleration
fields are interchanged when going from the outside the cone to the inside of the cone, meaning that inside the cone the observer
sees the field from the deceleration first. Fig.~\ref{fig:tracktime3} illustrates the narrowing of the electric field as the observer
moves closer to the Cherenkov angle.

\begin{figure}
\includegraphics[width=0.75\textwidth]{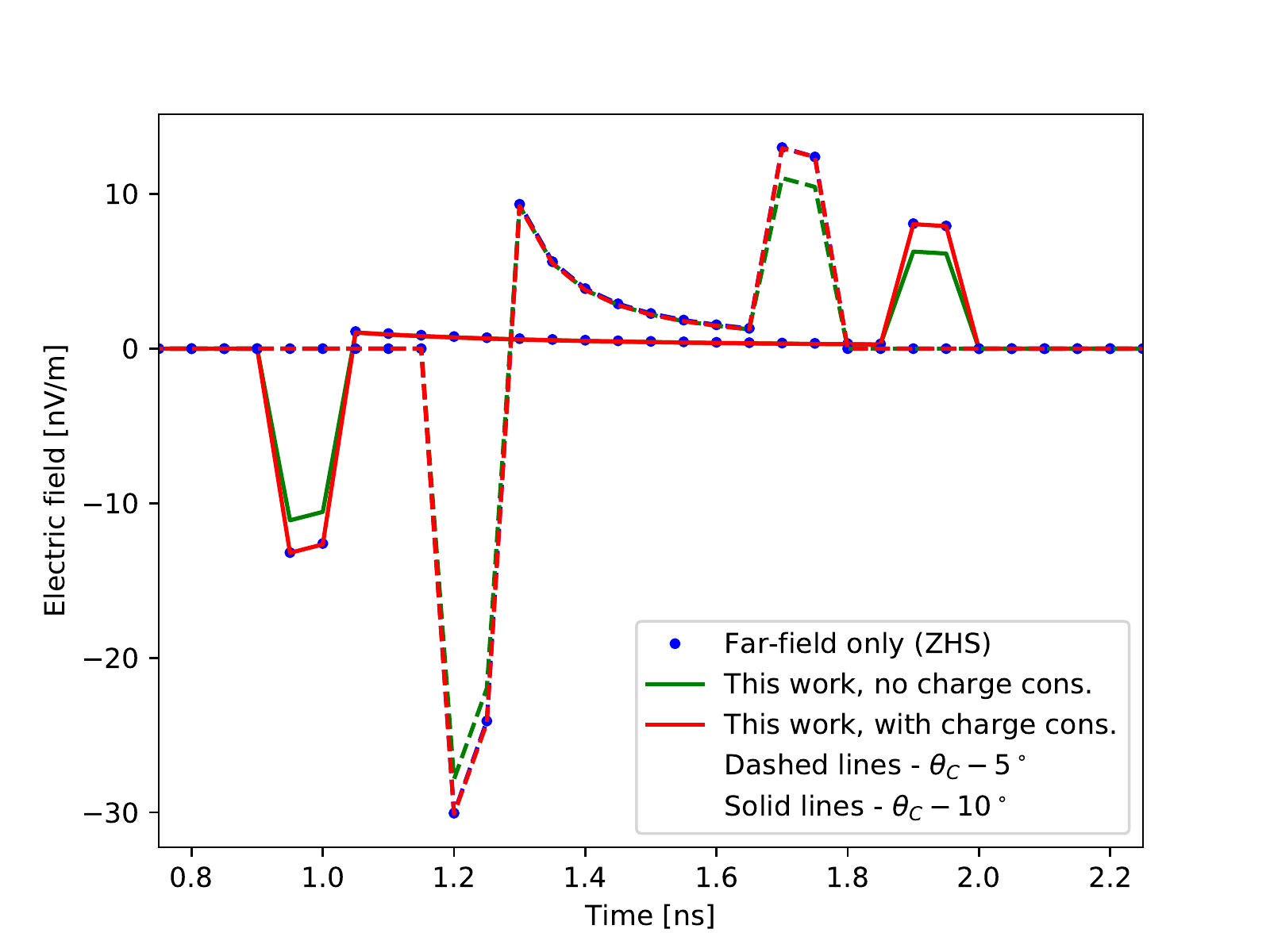}
\caption{Same as Fig.~\ref{fig:tracktime} but only the $x$ component of the field and with an observer at $\theta_C-10^\circ$
(solid lines) and another one at $\theta_C-5^\circ$ (dashed lines).}
\label{fig:tracktime3}
\end{figure}

Now that we have shown that our formula yields sensible results in time domain, we
will proceed with the study of its frequency domain counterpart.

\section{Electric field of a particle track in frequency domain}
\label{sec:freq}

We have two ways to know the emission of a particle track in frequency domain.
Either we transform Eq.~\eqref{eq:Etime2} or we solve Maxwell's equations in
frequency domain. We will adopt both approaches and check that they give, in
fact, the same equation.

\subsection{Transformation to frequency domain}

Eq.~\eqref{eq:Etime2} can be transformed to frequency domain. 
Defining the Fourier transform of a function $f(t)$ as
\begin{equation}
F(\omega) = \int_{-\infty}^{+\infty} \mathrm{d} t \ e^{i\omega t} f(t),
\end{equation}
and knowing the Fourier transform for a step function:
\begin{equation}
\int_{-\infty}^{+\infty} \mathrm{d} t \ e^{i\omega t} \Theta(t - t_0) = e^{i\omega t_0}
\left[ \frac{i}{\omega} + \frac{\delta(\omega)}{2} \right],
\end{equation}
we find that
\begin{equation}
-\frac{q}{4\pi\epsilon} \int_{-\infty}^{+\infty} \mathrm{d} t \frac{\hatR_1}{R_1^2} \Theta(t - t_1 - R_1/c_n) =
-\frac{q}{4\pi\epsilon} \frac{\hatR_1}{R_1^2} e^{i\omega t_1} e^{ikR_1}
\left[ \frac{i}{\omega} + \frac{\delta(\omega)}{2} \right],
\label{eq:trans1}
\end{equation}
with $k = \omega/c_n$ and
\begin{equation}
-\frac{q}{4\pi\epsilon c_n} \int \mathrm{d} t \ e^{i\omega t} \frac{\hatR_1}{R_1^2} \delta(t-t_1-R_1/c_n)
= -\frac{q}{4\pi\epsilon c_n} e^{i\omega t_1} e^{ikR_1} \frac{\hatR_1}{R_1^2} .
\label{eq:trans2}
\end{equation}
The two terms for the deceleration can be written following the same steps. Taking the
first term of the second line in Eq.~\eqref{eq:Etime2} and transforming it to frequency,
supposing that $\theta > \theta_C$
\begin{eqnarray}
\frac{q}{4\pi\epsilon} \int_{-\infty}^{+\infty}  \mathrm{d}t \, e^{i\omega t}
\left[ \frac{\hatR}{\kappa R^2} (\rect)\right]\ret =
\frac{q}{4\pi\epsilon} \int_{t_1+R_1/c_n}^{t_2+R_2/c_n} \mathrm{d}t \, e^{i\omega t}
\left[\frac{\hatR}{\kappa R^2}\right]\ret \nonumber \\
= \frac{q}{4\pi\epsilon} \int_{t_1}^{t_2} \mathrm{d}t' \, e^{i\omega t'} e^{i k R(t')}
\frac{\hatR}{\kappa R^2} \frac{\mathrm{d} t}{\mathrm{d} t'} =
\frac{q}{4\pi\epsilon} \int_{t_1}^{t_2} \mathrm{d}t' \, e^{i\omega t'} e^{i k R(t')}
\frac{\hatR}{ R^2}
\label{eq:transplus}
\end{eqnarray}
where we have made the change of variable $t \rightarrow t'$. Taking a geometry in which
$\theta < \theta_C$ instead, yields
\begin{equation}
\frac{q}{4\pi\epsilon} \int_{t_2+R_2/c_n}^{t_1+R_1/c_n} \mathrm{d}t \, e^{i\omega t}
\left[\frac{\hatR}{\kappa R^2}\right]\ret = 
\frac{q}{4\pi\epsilon} \int_{t_2}^{t_1} \mathrm{d}t' \, e^{i\omega t'} e^{i k R(t')}
\frac{\hatR}{\kappa R^2} (-\kappa) =
\frac{q}{4\pi\epsilon} \int_{t_1}^{t_2} \mathrm{d}t' \, e^{i\omega t'} e^{i k R(t')}
\frac{\hatR}{ R^2},
\label{eq:transmin}
\end{equation}
which is exactly the same result as Eq.~\eqref{eq:transplus}. In fact, if we suppose that
for a point along the track, $\theta = \theta_C$, and we also suppose that the arrival time of
the shockwave to the observer is $t_{C,\mathrm{obs}} = t_C + R_C/c_n$,
the transform is
\begin{eqnarray}
\frac{q}{4\pi\epsilon} \left[ \int_{t_{C,\mathrm{obs}}}^{t_1+R_1/c_n}
\mathrm{d}t \, e^{i\omega t} \left[ \frac{\hatR}{\kappa R^2} \right]\ret +
\int_{t_{C,\mathrm{obs}}}^{t_1+R_1/c_n}
\mathrm{d}t \, e^{i\omega t} \left[\frac{\hatR}{\kappa R^2} \right]\ret
\right]
= \nonumber \\
\frac{q}{4\pi\epsilon} \left[ \int_{t_1}^{t_C} \mathrm{d}t' \, e^{i\omega t'} e^{i k R(t')}
\frac{\hatR}{ R^2} +
\int_{t_C}^{t_2} \mathrm{d}t' \, e^{i\omega t'} e^{i k R(t')}
\frac{\hatR}{ R^2} \right] =
\frac{q}{4\pi\epsilon} \int_{t_1}^{t_2} \mathrm{d}t' \, e^{i\omega t'} e^{i k R(t')}
\frac{\hatR}{ R^2},
\label{eq:transcher}
\end{eqnarray}
exactly the same that Eqs.~\eqref{eq:transplus} and~\eqref{eq:transmin}. We can transform
directly the rest of the terms of the second line in Eq.~\eqref{eq:Etime2} using the derivative
property of the Fourier transform:
\begin{equation}
\int \mathrm{d}t \, e^{i\omega t} \frac{\partial f\xt}{\partial t}=
-i \omega \int \mathrm{d}t \, e^{i\omega t} f\xt,
\end{equation}
since for our function, $\lim_{t\rightarrow \pm \infty} f\xt = 0$.
Together with Eq.~\eqref{eq:transcher} we write
the following exact field for a track in frequency.
\begin{eqnarray}
\mathbf{E}\xw = \frac{q}{4\pi\epsilon} \Bigg\{ 
& - &\frac{\hatR_1}{R_1} e^{i\omega t_1} e^{i k R_1} \left[ \frac{i}{\omega R_1} + \frac{1}{c_n} \right]
+ \int_{t_1}^{t_2} \mathrm{d}t' \, e^{i\omega t'} e^{ikR} \left[
\frac{\hatR}{R^2} - \frac{i\omega}{c_n} \frac{\hatR}{R}
+ \frac{i\omega}{c_n^2} \frac{\mathbf{v}}{R} \right] \nonumber \\
& + & \frac{\hatR_2}{R_2} e^{i\omega t_2} e^{i k R_2} \left[ \frac{i}{\omega R_2} + \frac{1}{c_n} \right]
\Bigg\}
\label{eq:Efreq}
\end{eqnarray}
We have dropped the terms involving $\delta(\omega)$, since we are not interested in
the contribution at zero frequency.

\subsection{Derivation of the formula in frequency domain}

Eq.~\eqref{eq:Efreq} can be derived directly in frequency domain. Let us assume, without
loss of generality, a track that starts moving at $t_1$ at $z = z_1$ with a velocity
$\mathbf{v} = v \hat{z}$. The potentials in frequency domain can be written as
\begin{equation}
\Phi\xw = \frac{1}{4\pi\epsilon} \intx ' \, 
\frac{e^{ik |\xdiff|}}{|\xdiff|} \rho\xpw
\label{eq:phifreq}
\end{equation}
\begin{equation}
\mathbf{A}\xw = \frac{\mu_0}{4\pi} \intx ' \, 
\frac{e^{ik |\xdiff|}}{|\xdiff|} \mathbf{J}\xpw
\label{eq:afreq}
\end{equation}
It should be noted that the changes $\epsilon \rightarrow \epsilon(\omega)$
and $\mu_0 \rightarrow \mu(\omega)$ are always possible, so we can always
generalise our final result for a frequency-dependent dielectric and magnetic medium.
The charge density and the current density are steadily written as
\begin{eqnarray}
\rho\xpw  =  q\int_{-\infty}^{\infty} \mathrm{d}t' e^{i\omega t'}
& [ & \deltaprim{1}\thetaprim{1} +
\delta^{3}(\mathbf{x}' - \mathbf{x}_p(t')) (\rect) \nonumber \\
& + & \deltaprim{2}\thetaprim{2} ],
\label{eq:rhofreq}
\end{eqnarray}
\begin{equation}
\mathbf{J}\xpw = q v\hat{z} \int_{-\infty}^{\infty} \mathrm{d}t' e^{i\omega t'}
\delta^{3}(\mathbf{x}' - \mathbf{x}_p(t')) (\rect),
\label{eq:jfreq}
\end{equation}
with $\mathbf{x}_p(t') = \mathbf{x}_1 + \mathbf{v}(t'-t_1) = z_1\hat{z} + v\hat{z}(t'-t_1)$.
Using Eq.~\eqref{eq:rhofreq}, we split Eq.~\eqref{eq:phifreq} into three terms,
\begin{equation}
\Phi\xw =  \frac{q}{4\pi\epsilon} \left[  -\int_{t_1}^{\infty} \mathrm{d}t' \, \frac{e^{ikR_1}}{R_1} e^{i\omega t'}
+ \int_{t_2}^{\infty} \mathrm{d}t' \, \frac{e^{ikR_2}}{R_2} e^{i\omega t'}
+ \int_{t_1}^{t_2} \mathrm{d}t' \, \frac{e^{ikR}}{R} e^{i\omega t'}
\right],
\end{equation}
since $R = |\mathbf{x}_p(t') - \mathbf{x}|$ after integrating the three-dimensional Dirac deltas. Equivalently,
\begin{equation}
\Phi\xw = \Phi_1 + \Phi_2 + \Phi_p.
\end{equation}
Eq.~\eqref{eq:afreq} can be written as
\begin{equation}
\mathbf{A}\xw = \frac{\mu_0}{4\pi} \int_{t_1}^{t_2} \mathrm{d}t' \, \frac{e^{ikR}}{R} e^{i\omega t'} \mathbf{v}
\label{eq:afreq2}
\end{equation}
Due to the cylindrical symmetry of our problem, we can calculate the radial component of the
field using $\Phi$ alone.
\begin{equation}
E_\rho = -\frac{\partial}{\partial \rho} (\Phi) = -\frac{\partial}{\partial \rho} (\Phi_1 + \Phi_2 + \Phi_p)
\end{equation}
Using that
\begin{equation}
\frac{\partial}{\partial x_i}\left(
\frac{e^{ik|\xdiff|}}{|\xdiff|} \right) = 
\frac{e^{ik|\xdiff|}}{|\xdiff|} \left[
ik - \frac{1}{|\xdiff|} \right] \frac{\hat{x_i}\cdot(\xdiff)}{|\xdiff|},
\end{equation}
where $x_i = x, y, z$ but it can also be $\rho$, the radial coordinate. Let us write
\begin{equation}
-\frac{\partial}{\partial \rho}\Phi_p = 
\frac{q}{4\pi\epsilon}\int_{t_1}^{t_2} \mathrm{d}t' \, \frac{e^{ikR}}{R} e^{i\omega t'}
\left[ \frac{1}{R} - \frac{i\omega}{c_n} \right] \hatR\cdot\hat{\rho}.
\label{eq:phip}
\end{equation}
Using the transform of the step function, we find the derivatives of $\Phi_1$ and $\Phi_2$.
\begin{equation}
-\frac{\partial}{\partial \rho}\Phi_1 =
-\frac{q}{4\pi\epsilon}\int_{t_1}^{\infty} \mathrm{d}t' \, \frac{e^{ikR_1}}{R_1} e^{i\omega t'}
\left[ \frac{1}{R_1} - \frac{i\omega}{c_n} \right] \hatR_1\cdot\hat{\rho} =
-\frac{q}{4\pi\epsilon} \frac{e^{ikR_1}}{R_1} e^{i\omega t_1}\frac{i}{\omega}
\left[ \frac{1}{R_1} - \frac{i\omega}{c_n} \right] \hatR_1\cdot\hat{\rho},
\label{eq:phi1}
\end{equation}
\begin{equation}
-\frac{\partial}{\partial \rho}\Phi_2 =
\frac{q}{4\pi\epsilon} \frac{e^{ikR_2}}{R_2} e^{i\omega t_2}\frac{i}{\omega}
\left[ \frac{1}{R_2} - \frac{i\omega}{c_n} \right] \hatR_2\cdot\hat{\rho}.
\label{eq:phi2}
\end{equation}
We have dropped the terms with $\delta(\omega)$, once again.
Formally, the derivatives with respect to $z$ are the same but making the replacement
$\hat{\rho} \rightarrow \hat{z}$. This fact, along with the identity
\begin{equation}
\hatR\cdot\hat{\rho} + \hatR\cdot\hat{z} = \hatR 
\end{equation}
and also that $\mu_0 = 1/(\epsilon c_n^2)$,
helps us to write the total electric field. Recalling Eqs.~\eqref{eq:afreq2}, \eqref{eq:phip},
\eqref{eq:phi1} and~\eqref{eq:phi2}:
\begin{eqnarray}
\mathbf{E}\xw & = & -\nabla\Phi + i\omega\mathbf{A} = \nonumber \\
 \frac{q}{4\pi\epsilon} \Bigg\{
& - & \frac{\hatR_1}{R_1} e^{i\omega t_1} e^{i k R_1} \left[ \frac{i}{\omega R_1} + \frac{1}{c_n} \right]
+ \frac{\hatR_2}{R_2} e^{i\omega t_2} e^{i k R_2} \left[ \frac{i}{\omega R_2} + \frac{1}{c_n} \right]
\nonumber \\
& + & \int_{t_1}^{t_2} \mathrm{d}t' \, \frac{e^{ikR}}{R} e^{i\omega t'}
\left[ \frac{1}{R} - \frac{i\omega}{c_n} \right] \hatR +
\int_{t_1}^{t_2} \mathrm{d}t' \, \frac{e^{ikR}}{R} e^{i\omega t'} \frac{i\omega}{c_n^2}\frac{\mathbf{v}}{R}
 \Bigg\},
\label{eq:Efreq2}
\end{eqnarray}
which is exactly the same as Eq.~\eqref{eq:Efreq}. Since we can always find a frame where the
track lies along the $z$ axis, and Eq.~\eqref{eq:Efreq2} is the same as Eq.~\eqref{eq:Efreq}, that
means that Eq.~\eqref{eq:Etime2} contains the exact field of a particle track at all frequencies.

\subsection{Calculation of frequency-domain field and comparison with far-field (ZHS) formula}

Another frequency-domain equation for the same problem of a track traveling
along the $z$ axis can be found in \cite{prd87}. The radial ($\rho$) and vertical
($z$) components of the electric field can be written in this work's notation as follows:

\begin{equation}
E_\rho\xw = i \frac{qv}{\omega}\frac{1}{4\pi\epsilon}
\int_{t_1}^{t_2} \mathrm{d}t' \, e^{i\omega t'} \frac{e^{ikR}}{R}
\sin\theta\cos\theta \left[
b \left( b - \frac{1}{R} \right) + \frac{1}{R^2} 
\right],
\label{eq:Eoldrho}
\end{equation}
and
\begin{eqnarray}
E_z\xw & = & i\frac{qv}{\omega}\frac{1}{4\pi\epsilon}
\int_{t_1}^{t_2} \mathrm{d}t' \, e^{i\omega t'} \frac{e^{ikR}}{R}
\left[
b^2 \cos^2\theta + \frac{\cos^2\theta}{R^2} -
\frac{b}{R}\left(\cos^2\theta - 1 \right)
\right] \nonumber \\
& + & i\omega \frac{\mu_0}{4\pi} qv
\int_{t_1}^{t_2} \mathrm{d}t' \, e^{i\omega t'} \frac{e^{ikR}}{R},
\label{eq:Eoldz}
\end{eqnarray}
where $b(t') = ik - \frac{1}{R}$. Eqs.~\eqref{eq:Eoldrho} and~\eqref{eq:Eoldz} have been derived
following the same premises as Eq.~\eqref{eq:Efreq}, but circumventing the problem of
charge conservation by explicitely using the Lorenz gauge condition in frequency domain. This way,
it is possible to calculate the scalar potential $\Phi$ by means of the vector potential $\mathbf{A}$, that is
\begin{equation}
\mathbf{A}\xw = i\epsilon\mu_0\omega\,\Phi\xw,
\end{equation}
allowing us to arrive at a correct field without having to write a correct charge density. In exchange,
we arrive at a more cumbersome electric field (Eqs.~\eqref{eq:Eoldrho} and~\eqref{eq:Eoldz}) than
in the present work (Eq.~\eqref{eq:Efreq}), although both expressions are equivalent at non-zero frequency.

We show an example of this equivalence in Fig.~\ref{fig:trackfreq}, where both approaches give the same
numerical result for an observer placed at the Cherenkov angle of a track. As a consequence, Eq.~\eqref{eq:Efreq}
contains the exact field of a track, including the Cherenkov shock wave \cite{afanasiev}.
In particular, this means that in far-field regime, Eq.~\eqref{eq:Efreq} agrees with the
ZHS algorithm for large frequencies ($kR \gg 1$), which is shown in Fig.~\ref{fig:trackfreq}.

\begin{figure}
\includegraphics[width=0.75\textwidth]{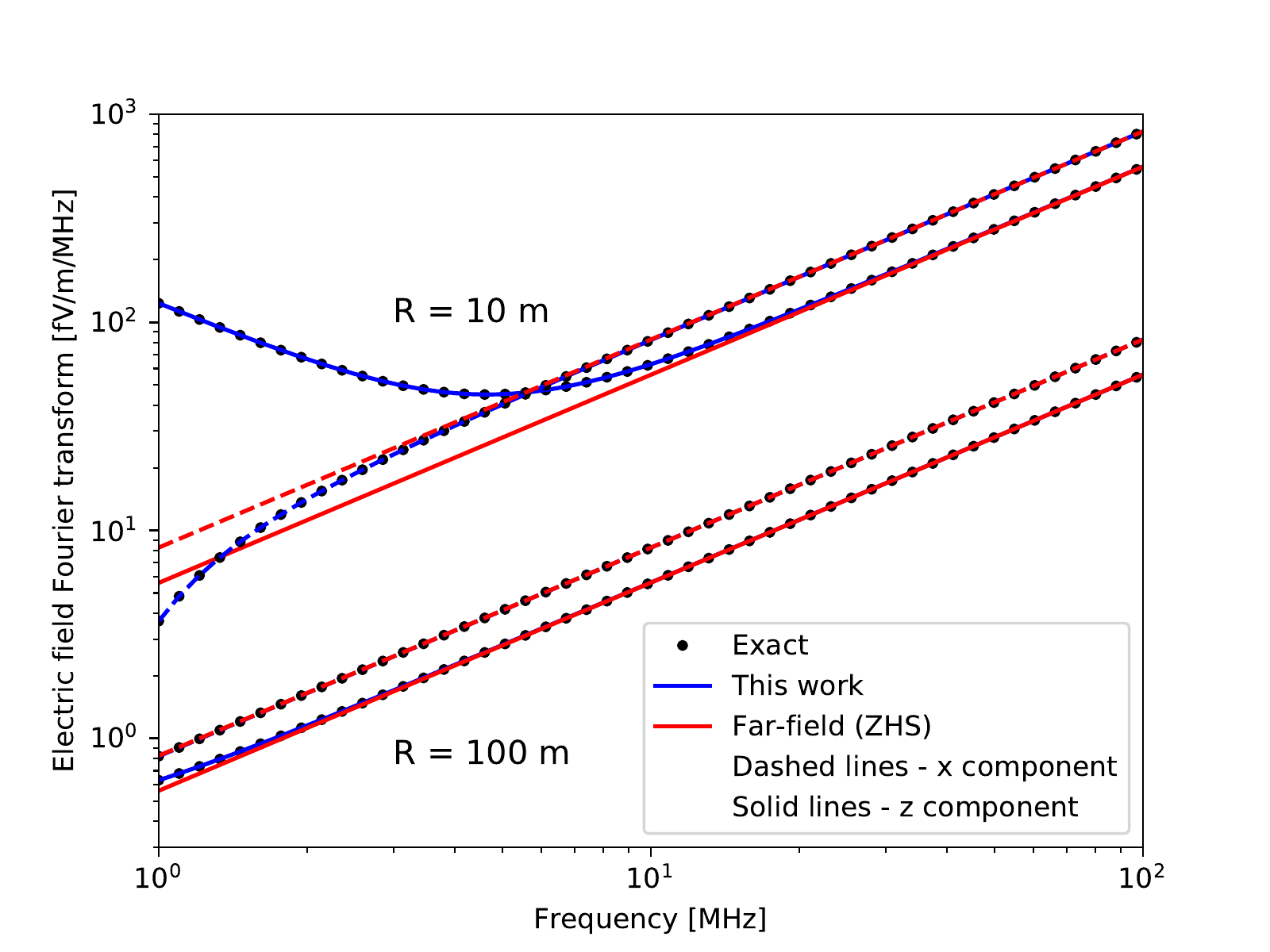}
\caption{Fourier transform of the electric field ($x$ and $z$ components) as a function of frequency for a
$1.2$ m long track traveling in deep Antarctic ice ($n = 1.78$) along the $z$ axis at a speed $v \approx c$. 
The observer is placed at the Cherenkov angle, at distance of $R = 10$ m and $R = 100$ m.
The exact field from
Eqs.~\eqref{eq:Eoldrho} and~\eqref{eq:Eoldz} (black points) is plotted, along with the field from Eq.~\eqref{eq:Efreq} (blue lines) 
and the field calculated with the ZHS algorithm (red lines). The first two approaches are indistinguishable,
and the third one agrees with them in the far-field regime.}
\label{fig:trackfreq}
\end{figure}

\section{Implementation of the time-domain formula in the SELFAS Monte Carlo code}
\label{sec:mc}

Once it has been shown that Eq.~\eqref{eq:Etime2} is an adequate formula for the description
of the electric field created by a particle track, we can embed this formula in an air shower
simulation code in order to calculate the electric field generated by an extensive air shower.
We have chosen the SELFAS \cite{selfas} MC code. SELFAS uses a \emph{hybrid} approach for the
simulation -- a given number of particles and their properties are sampled from probability distributions
(longitudinal and lateral positions, velocity...) and then each particle is subject to interactions
and microscopically followed for an atmospheric depth of $15$ g$/$cm${}^{2}$. The combination
of macroscopic particle distibutions and microscopic treatment of each individual particle constitutes
what we call the hybrid approach. Currently, SELFAS samples the particles using 
the longitudinal shower profile
given by CONEX \cite{conex}, a lightweight code that produces results compatible with the full
MC CORSIKA. SELFAS has been recently upgraded with a state-of-the-art treatment of the
atmosphere with its inclusion of the Global Data Assimilation System (GDAS) \cite{florian,benoiticrc}.

Before we implement Eq.~\eqref{eq:Etime2} in the code, special attention has to be paid to the
Coulomb terms of the static particles left behind, since they are not detectable in an EAS. Firstly,
because the physical situation is not the same. If a random track inside the shower is picked,
there is a high probability that this track is the continuation of an ongoing particle trajectory and
not the beginning, meaning there is no opposite charge left behind and its static field can be ignored.
Analogously, a randomly chosen track will likely not be the end of the true particle trajectory, so we will
not suppose that the particle rests at the end of the track and ignore its Coulomb field. Secondly,
if there are static charged particles left behind, the medium itself will change in order to regain
electrostatic equilibrium and minimise the static electric fields. Because of these physical reasons,
we will not compute the Coulomb terms $\frac{\hatR_{1,2}}{R_{1,2}^2} \Theta(t - t_{1,2} - R_{1,2}/c_n)$.

Another important matter is what to do with the particles that reach the ground, since our calculation
is valid for a single medium. When a track reaches ground, its field can be calculated as the sum of
the field of a track that stops just above the ground and another one accelerating at the same time
just below the ground, and making their distance as small as possible. This approach explains
transition radiation \cite{motloch}. With this scheme in mind, we can separate the total field into
three different contributions:
\begin{itemize}
\item The direct field coming from the track above the surface (as if there were no boundary). This
contribution can be calculated, at all frequencies, with Eq.~\eqref{eq:Etime2}.
\item The field created by the response of the interface (in this case, the ground) to the field
emitted by the track above the surface. If the track lies in the far-field with respect to the ground,
and also the antenna lies in the far-field with respect to the ground, that is, if the distances to
the ground are much larger than the observation wavelength, we can use geometrical optics and
the standard Fresnel coefficients. If the antenna is close to the ground but the track
is in the far-field with respect to the ground, a possible workaround is
to use the reciprocity theorem \cite{balanis,frank}, which allows us to know the antenna voltage
with the far-field antenna pattern and the direct field of the track.

In our case, since we are interested in frequencies around the MHz ($\lambda \sim 300$ m),
and there is a non-negligible number of particles arriving at ground, we cannot use geometrical
optics. Moreover, the antennas are placed close to the ground in most radio EAS experiments
($9$ m high in the case of EXTASIS \cite{icrcarxiv}). 
When the emitter and receiver are in the near field of
the ground, what we have is a wave that is a mixture of a surface wave and a reflected
(in the sense of geometrical optics) wave \cite{surface}, which can be more similar to the
former or the latter depending on the actual physical configuration. The calculation of such a field lies
outside of the scope of the present paper, so as a first order approximation we will not calculate
this contribution. Nevertheless, the importance of this contribution constitutes an interesting
study that we are currently carrying out.
\item The contribution of the underground track will be ignored. Because of the larger attenuation
losses of radio waves inside soil \cite{soil}, we expect this contribution to be much smaller than
the surface wave contribution.
\end{itemize}
To sum up with, we will calculate the direct contribution from the parts of the track above the ground.
This is equivalent to suddenly stopping the particles upon their arrival to the ground, which seams
reasonable, since this electrons are in fact beta radiation with a mean speed of $\sim 0.3c$ and 
they are stopped in $<1$ cm of soil.

\section{Simulations for EAS and comparison with the far-field emission}
\label{sec:sim}

We have simulated a proton-induced 1 EeV shower, with a zenith angle of $30^\circ$
and coming from the east ($\phi = 0^\circ$ marks the east and $\phi = 90^\circ$ the north), 
using $2\times 10^8$ particles. The ground lies at an altitude of $1400$ m, as it is the
case for the Pierre Auger Observatory.
Several antennas have been placed north of the shower core,
at different distances from it. Since Eq.~\eqref{eq:Etime2} is equivalent to the ZHS formula
in the far-field, we will focus on the low frequency part of the electric field and filter our
resulting electric fields with a lowpass (or bandpass, depending on the case) sixth-order
Butterworth filter\footnote{A Butterworth filter has the important property of preserving
causality. An unphased step or box filter
%that is, transforming the trace to frequency, preserving
%only the frequencies we are interested in and multiplying the spectrum by zero elsewhere, 
will give an acausal behaviour. In particular, for a typical EAS signal, it will produce an unphysical
ringing with a non-zero field well before the actual arrival time of the wave
at the observing location.}
%This is undesirable not only from a theoretical point of view, but also from a practical one, if
%we want to study arrival times, for instance.}. 
We show in Fig.~\ref{fig:trace_v} the resulting
vertical component of the electric field (filtered at frequencies $< 5$ MHz) for 
one observer at $200$ m east of the shower core and another at $500$ m. 
The electric fields have been calculated with the far-field (ZHS) and complete approach
(Eq.~\eqref{eq:Etime2}) using the very same shower.
It is worth noting
that at low frequency (Fig.~\ref{fig:trace_v}, top, Fig.~\ref{fig:trace_ew}) 
and at a distance of $200$ m, differences
between the far-field and the complete approach are quite important, whereas for $500$ m
both approaches are visually indistinguishable. This is also the case 
when filtering between $[20-80]$ MHz
for the observer at $200$ m (Fig.~\ref{fig:trace_v}, bottom). Another important feature found
is the existence of a pulse originated by the sudden deceleration of
the shower front, marked by the arrows in Fig.~\ref{fig:trace_v}. Note that this deceleration
is coherent only at low frequencies, since the pulse disappears filtering in the $[20-80]$ MHz band.
From now on, we will call this pulse the sudden death pulse (SDP) \cite{sdpicrc}.

In Fig.~\ref{fig:trace_ew} we show the geomagnetic component (EW), finding that at low frequencies,
the main geomagnetic pulse from the EAS is changed in a non-negligible way 
at $200$ m from the shower core.
A remarkable fact, because the altitude of the shower maximum being at several kilometers,
one could think that the far-field approach could suffice for computing near the shower
axis at $\sim$MHz frequencies, but this is not the case.

\begin{figure}
\includegraphics[width=0.6\textwidth]{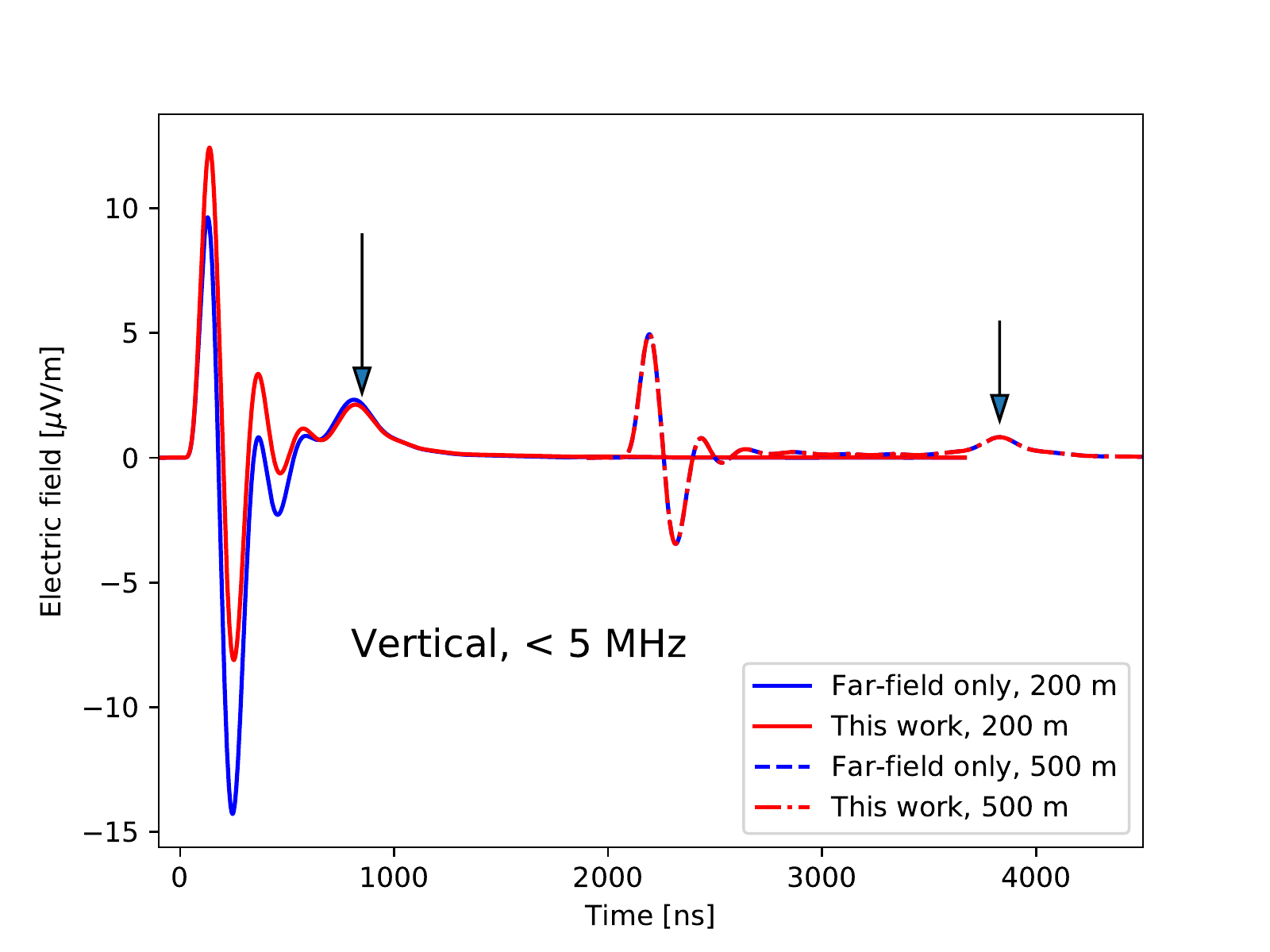}
\includegraphics[width=0.6\textwidth]{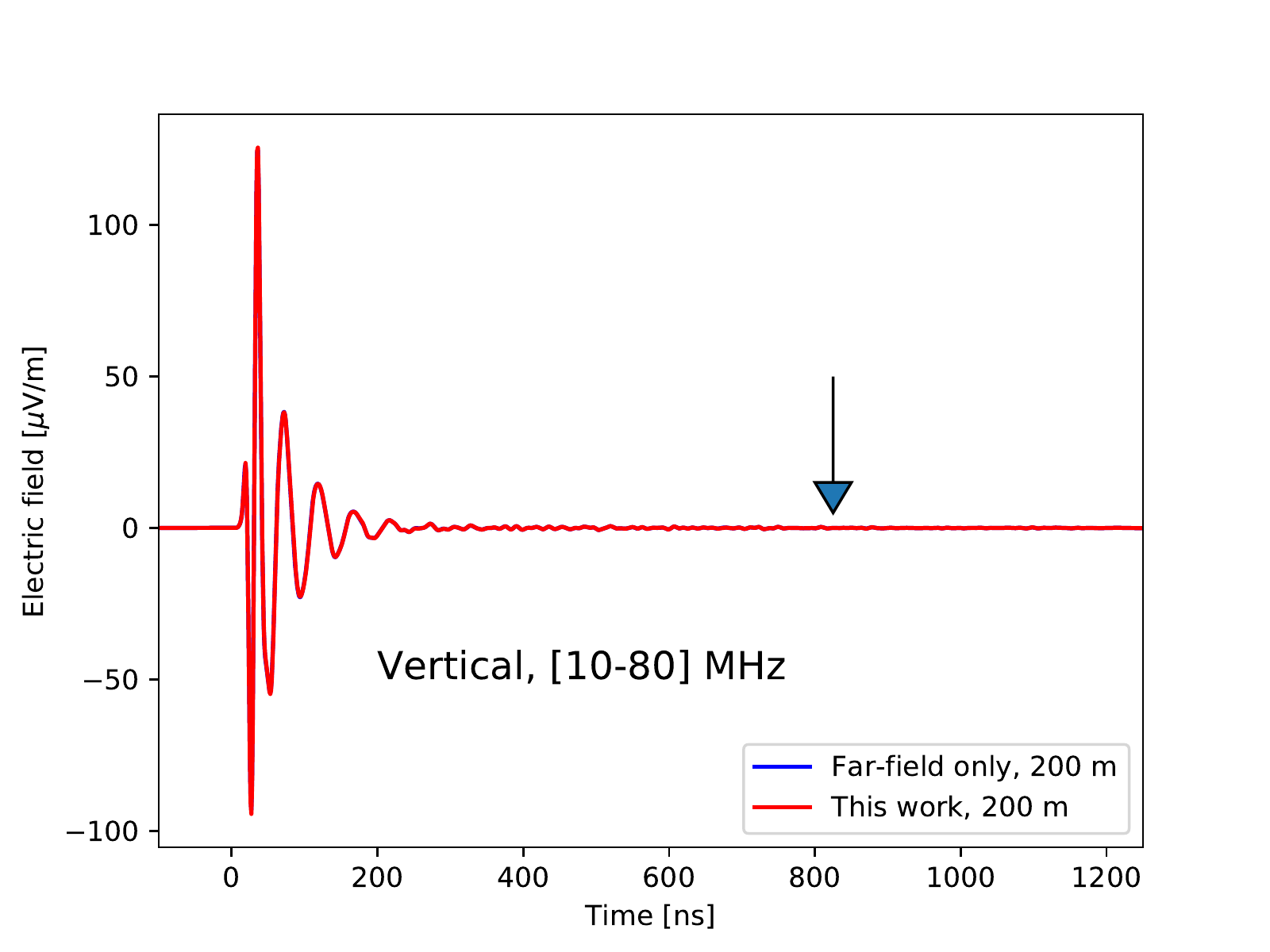}
\caption{Electric field as a function of time created by a $1$ EeV proton-induced shower with $30^\circ$ of
zenithal angle and coming from the East ($\phi = 0^\circ$). Times have been
arbitrarily offset. Traces have been numerically transformed to frequency, then filtered
with a sixth order low-pass Butterworth filter and transformed back to
time domain. This work's formula (red lines) and the far-field approximation (ZHS,
blue lines) are plotted.
Observers have been placed at $200$ m (solid lines) and $500$ m (dashed lines) 
east from the shower core.
The sudden death field (indicated by the arrows) is visible after the
principal pulse in each trace below $5$ MHz. Top: $5$ MHz low-pass filter. Bottom: $10-80$ MHz
band-pass filter. See text for details.}
\label{fig:trace_v}
\end{figure}

\begin{figure}
\includegraphics[width=0.6\textwidth]{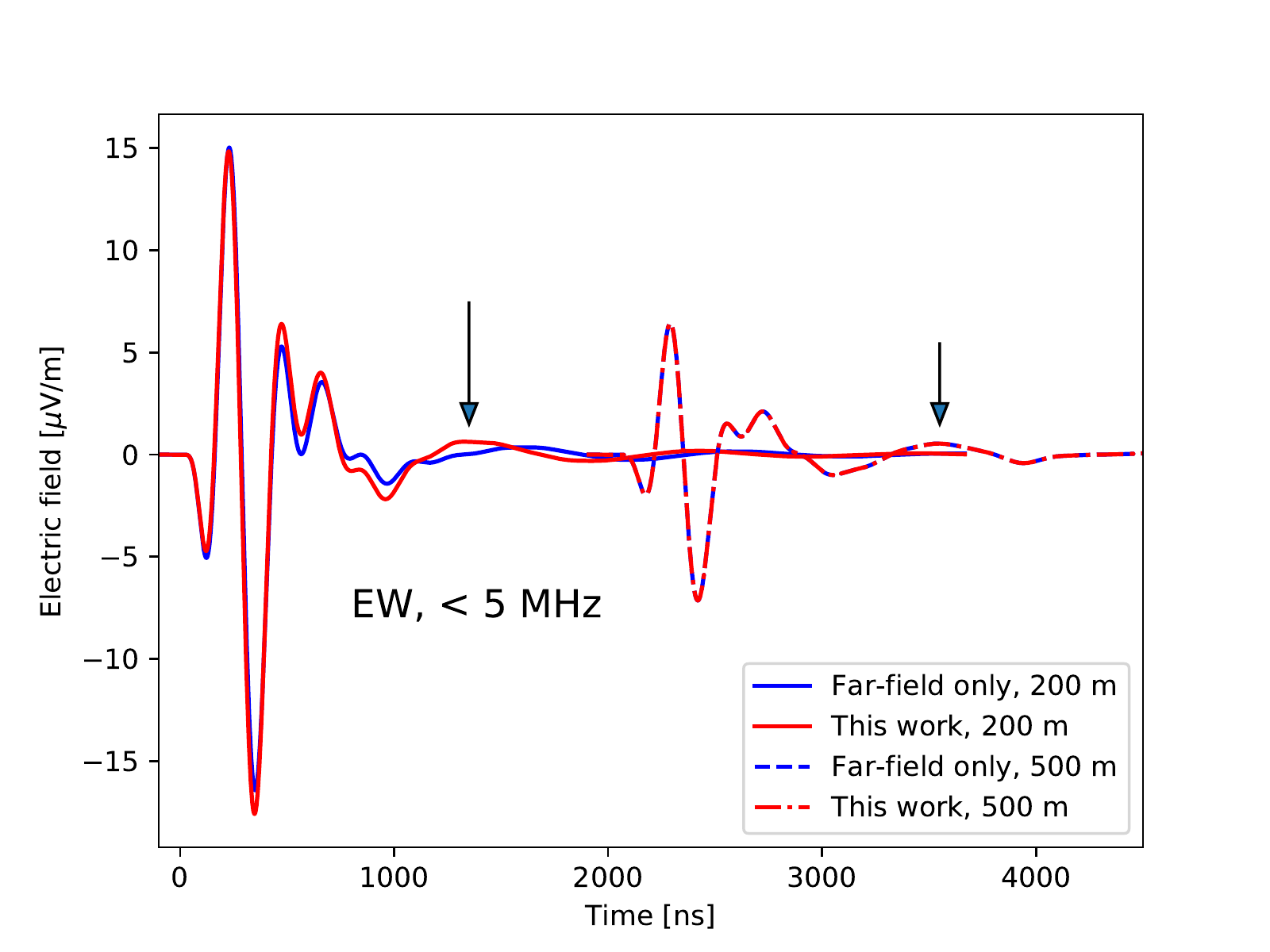}
\caption{Same as Fig.~\ref{fig:trace_v}, top, but for the EW polarization. See text for details.}
\label{fig:trace_ew}
\end{figure}

We show in Fig.~\ref{fig:spectrum} some examples of the spectrum amplitude of the electric field for
the far-field and complete approaches. For an observer at $500$ m, differences
between the two approaches are negligible, whereas at $200$ m and below $10$ MHz
differences are quite important. In order to quantify them, we show the relative difference
between the two approaches for the same
shower in Fig.~\ref{fig:error}, defined as the absolute value of the difference between the far-field 
and the exact formula divided by the amplitude of the exact formula (for a given polarization).
Below $10$ MHz, and for this given shower, both approaches differ considerably and one
should use the exact formula if a complete electric field is needed. The relative difference
falls with frequency until $\sim 10$ MHz, where it begins to be dominated by the incoherent emission of the particles of the lower part of the shower 
that are closer to the observer (located in the near field). 
The error lies around $\sim 1\%$ above $10$ MHz,
except for the vertical component at $50$ m from the shower core, where the discrepancy
between the two formulas is still large up to $\sim 50$ MHz.

\begin{figure}
\includegraphics[width=0.6\textwidth]{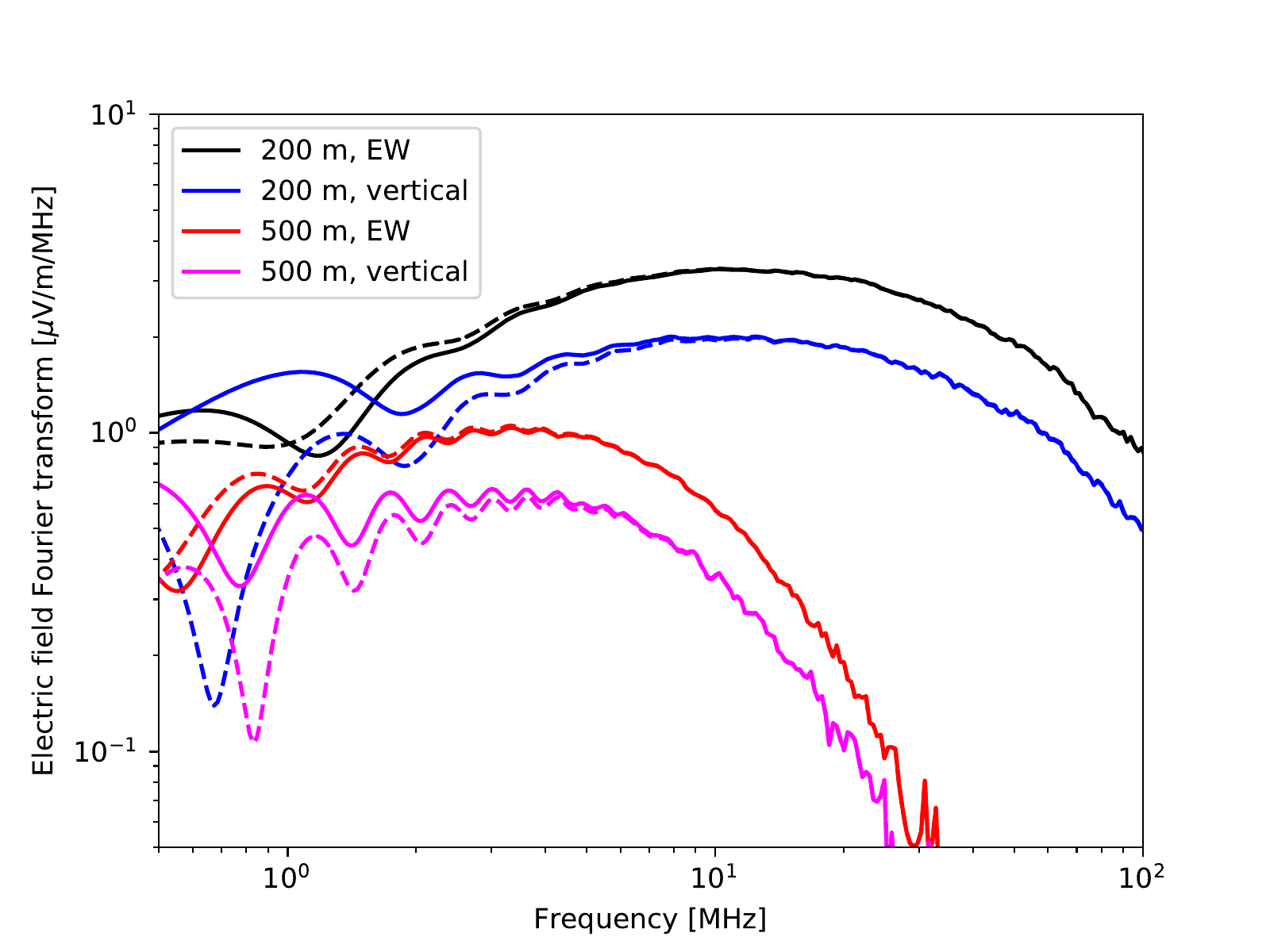}
\caption{Amplitude of the Fourier transform of the electric field as a function of frequency
created by a $1$ EeV proton-induced shower with $30^\circ$ of zenithal angle and
coming from the East ($\phi = 0^\circ$). Dashed lines indicate the result of the complete
formula, while solid lines portray the far-field calculation. We have plotted the East-West and
vertical polarizations for an observer at $200$ m and another at $500$ m north of
the shower core.}
\label{fig:spectrum}
\end{figure}

\begin{figure}
\includegraphics[width=0.6\textwidth]{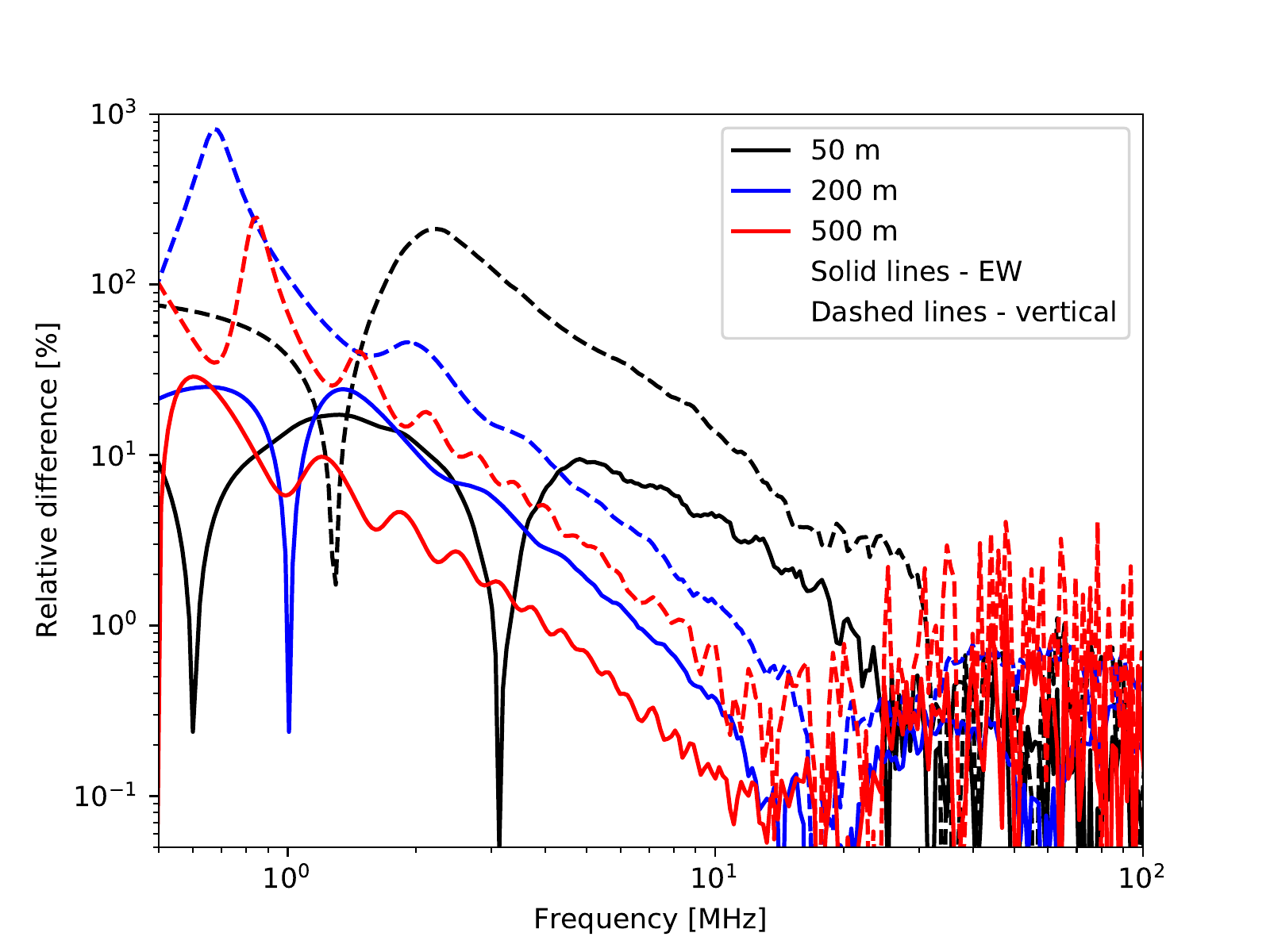}
\caption{Relative difference as a function of frequency 
between the exact solution and the far-field approximation (ZHS)
for the same shower of Fig.~\ref{fig:spectrum}. We have placed observers at $50$ (black), 
$200$ (blue) and $500$ m (red). Solid lines indicate EW polarization and dashed lines
vertical polarization.}
\label{fig:error}
\end{figure}

So as to elucidate whether we should use the exact formula instead of the far-field approximations
for most of today's radio experiments, that measure the electric field above $20$ MHz,
we have computed the relative difference as a function of the distance for a fixed frequency of $20$ MHz varying the zenithal angle of the shower. 
We have simulated, for each zenithal angle, ten different showers and calculated the
relative difference between the far-field and exact calculations.
We have plotted the mean and standard deviation of these differences 
in Fig.~\ref{fig:error_dist}, where we can conclude that at distances larger than $\sim 100$ m,
the error for the inspected showers is less than $1\%$ for the EW and NS components. 
However, the error in the vertical component for a vertical shower
(less than $20^\circ$) is quite large. For the depicted $0^\circ$ shower
(Fig.~\ref{fig:error_dist}, top left), even at 
$250$ m from the shower core, the mean error lies around $\sim 30\%$.
For the inspected $0$, $20$, $40$ degree showers, the error in the horizontal
polarizations at distances larger than $100$ m is of the order of $1\%$ or less,
but it does not decrease rapidly with the distance because of the shower particles
that lie closer to the ground. Because of this, the error comitted
when calculating the field from a $60^\circ$ shower
(Fig.~\ref{fig:error_dist}, bottom right), that presents a smaller number of particles
near the observer, is less than the error committed for vertical showers.
We have checked that for showers having a zenith angle smaller than
$40^\circ$ the error as a function of distance for $50$ MHz is similar to
the error at $20$ MHz.
The above facts hint that for most
current experiments measuring in the $[20-80]$ MHz band and only horizontal polarizations, 
the far-field approximation rests valid at distances larger than $100$ m from the shower core.
If, however, an antenna lies at a smaller distance, or if the vertical component is being
measured, one should consider using the
exact formula presented in this work. For low-frequency experiments,
the use of Eq.~\eqref{eq:Etime2} is imperative.

\begin{figure}
\includegraphics[width=0.49\textwidth]{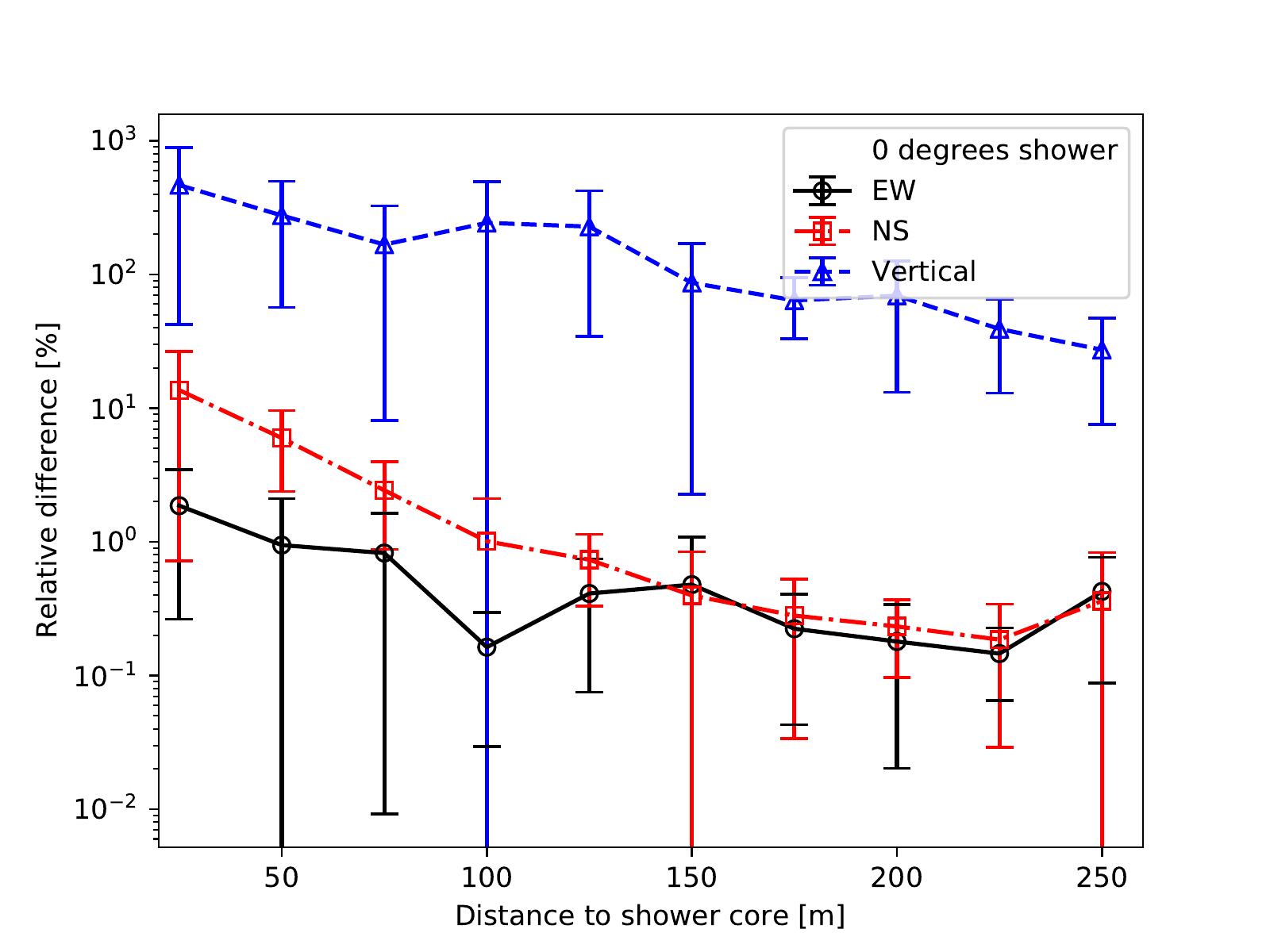}
\includegraphics[width=0.49\textwidth]{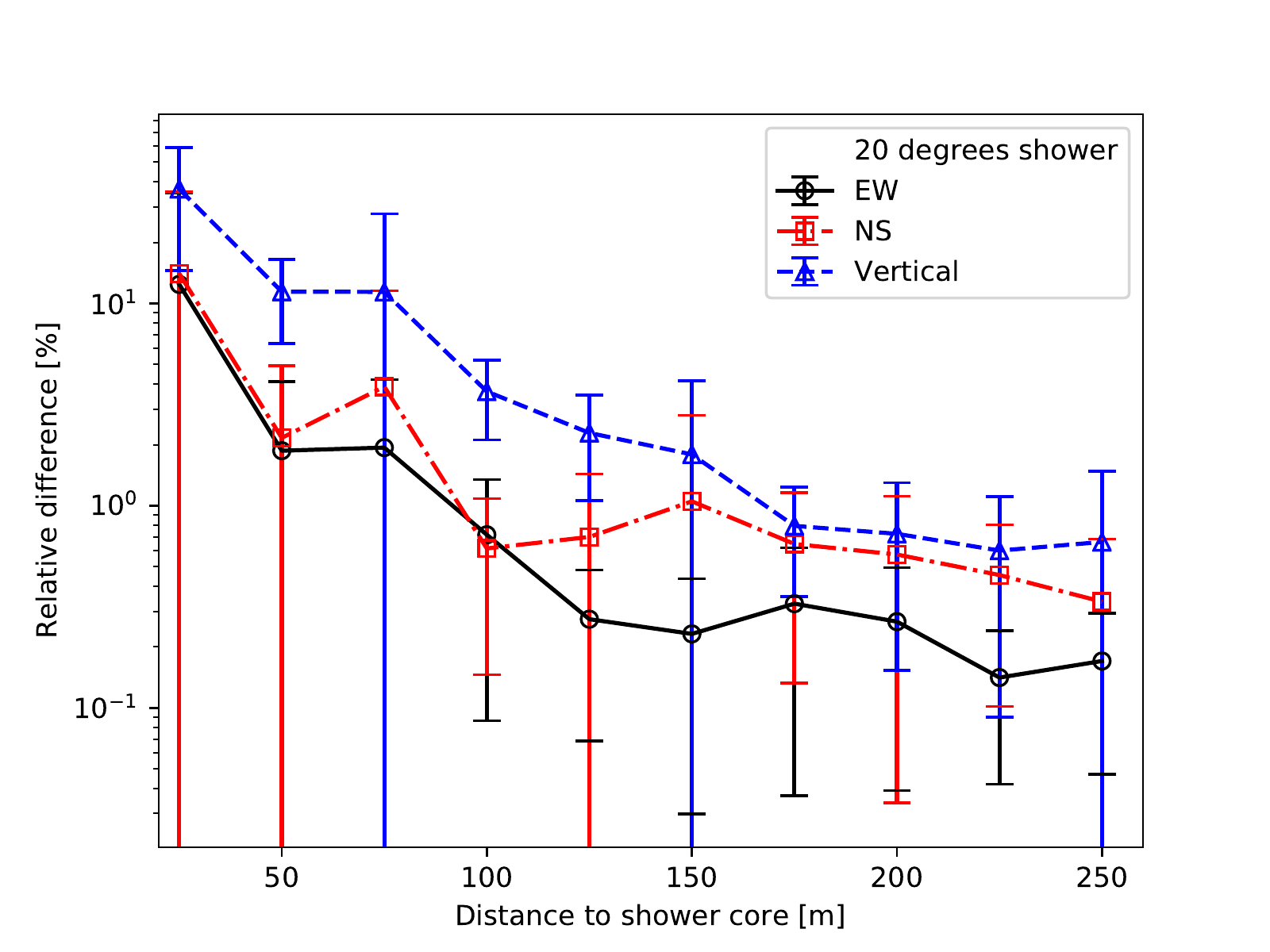}
\includegraphics[width=0.49\textwidth]{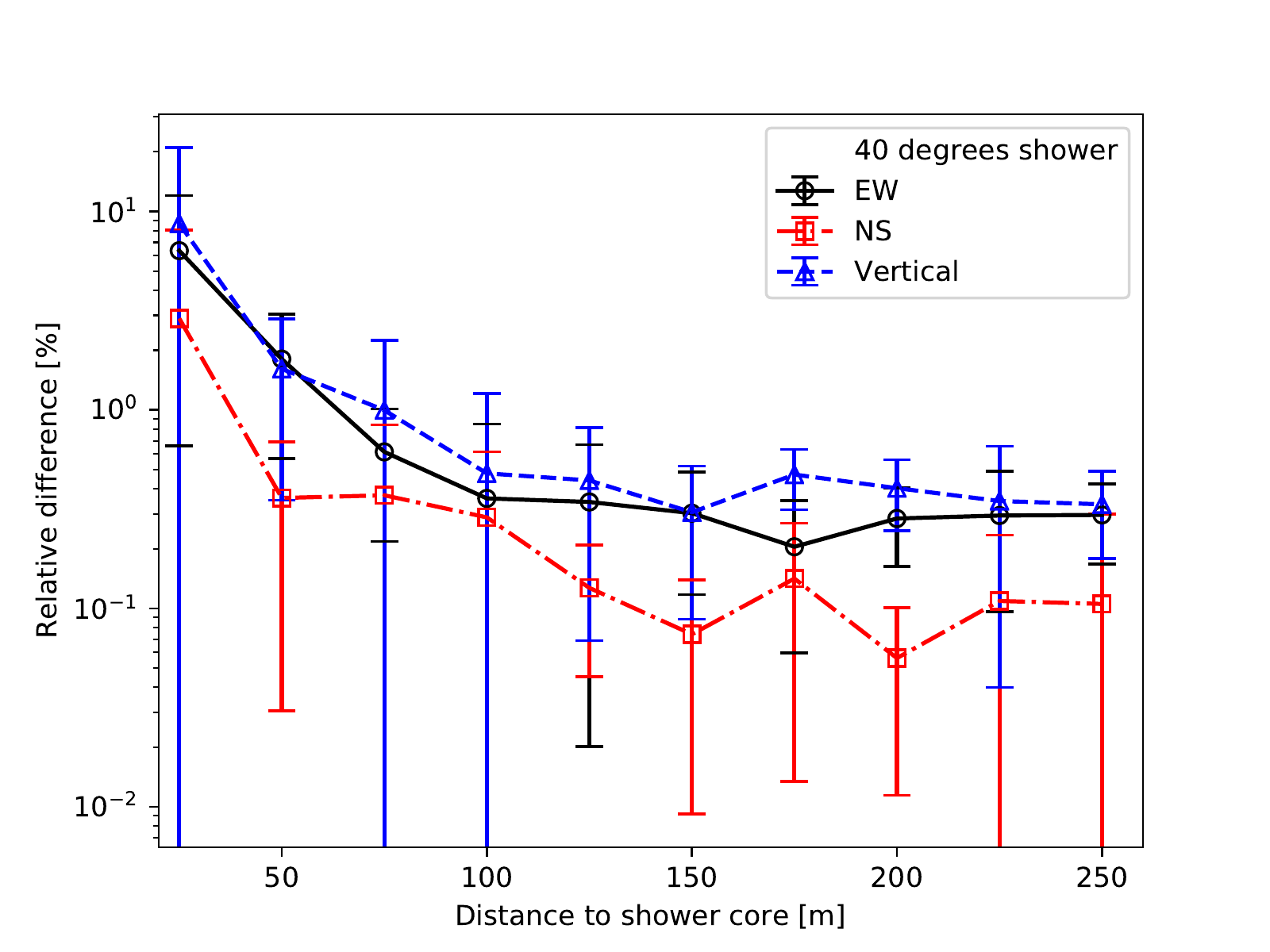}
\includegraphics[width=0.49\textwidth]{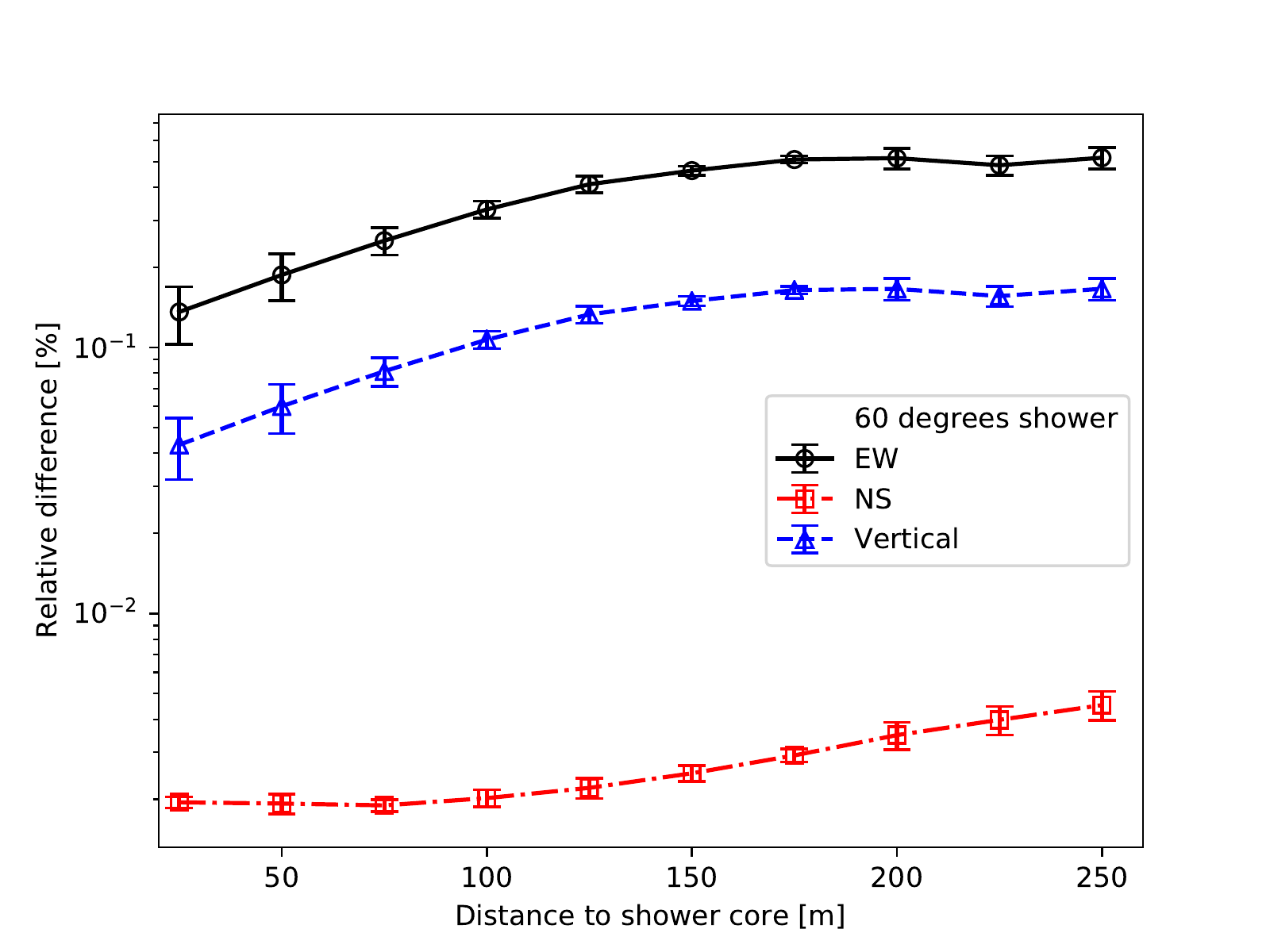}
\caption{Relative difference as a function of distance
between the exact solution and the far-field approximation (ZHS)
for the same shower of Fig.~\ref{fig:spectrum}. The relative differences have
been calculated for $1$ EeV showers at frequency of $20$ MHz and
averaged over 10 showers. The means of these differences
are represented by the marks, while the error bars indicate the standard
deviation. EW, NS and vertical polarizations are plotted as indicated in the
legend. 
Top left: $0^\circ$ zenith angle shower.
Top right: $20^\circ$ zenith angle shower.
Bottom left: $40^\circ$ zenith angle shower.
Bottom right: $60^\circ$ zenith angle shower.
}
\label{fig:error_dist}
\end{figure}

\section{Properties of the sudden death pulse}
\label{sec:sdp}
\subsection{Simple model for the sudden death pulse}
We have established that the SDP obtained in our simulations is a pulse 
created by the sudden deceleration of the shower front upon its arrival to the ground.
Because of the large extension of the shower front and its thickness that causes
the particles to arrive at different times, the pulse is only coherent at
low frequencies (Fig. \ref{fig:trace_v}). However, the NKG formula predicts that
even for relative old showers with an age of $s = 1.5$, around half of the particles will
be contained inside a distance to the shower axis less than 
the Moli\`{e}re radius, $\sim 100$ m. For a young shower
with $s = 1.2$, we will have more particles inside. The Moli\`{e}re radius is comparable
to the wavelenghts between 1 and 5 MHz, that is, from $300$ m to $60$ m, which inspires
us to postulate a simple model for the SDP, wherein all the particles travel parallel
to the shower axis and arrive to the shower core, but we still allow them to arrive at
different times in order to reproduce the shower longitudinal development. The model
implies that the bulk of the radiation will be produced when the part of the shower
with most particles suddenly decelerates, creating a maximum for the pulse of the form
\begin{equation}
\mathbf{E}_{\textrm{SDP,max}}\xt = \frac{Nq}{4\pi\epsilon R} \mathbf{v}_\perp
\delta(t - \frac{nR}{c}) =
\frac{Nq}{4\pi\epsilon R} (- \hatR\times(\hatR\times\mathbf{v}))
\delta(t - \frac{nR}{c})
\label{eq:sdp}
\end{equation}
where we have applied the far-field aproximation, and also 
that the core is located at the origin and the shower maximum
arrives to the ground at $t = 0$. $N$ is the number of particles arriving to the ground 
and $\mathbf{v}$ is the mean velocity of the ground particles, while
$\mathbf{v}_\perp = - \hatR\times(\hatR\times\mathbf{v})$ corresponds to the
projection of the velocity perpendicular to the line of sight that joins the shower
core and the observer.
Eq.~\eqref{eq:sdp} gives already quite some insight on the expected maximum
of the SDP. It implies that:
\begin{itemize}
\item the arrival time of the maximum of the pulse is linear with the distance to the
shower core, and the proportionality is precisely the inverse speed of light in the medium;
\item in the far-field, we expect a maximum amplitude with a dependence of $1/R$,
akin to a radiation field, and
\item the polarization of this pulse should lie along the direction
of the mean velocity of the ground particles
projected onto the direction perpendicular to the line of sight core-observer.
\end{itemize}

\subsection{Arrival time of the SDP as a function of the distance to the shower core}
We have simulated a $1$ EeV proton-induced shower of $30^\circ$ of zenithal angle
coming from the east and put several observers north of the shower core. After filtering
the traces with a $10$ MHz sixth-order Butterworth filter, we show the
SDP electric field as a function of time for different distances to the shower core
in Fig.~\ref{fig:tracesshift}, where it is apparent that the maxima present a displacement
proportional to the distance. Moreover, the amplitude seems to decrease with the
inverse of the distance. The traces in Fig.~\ref{fig:tracesshift} present an asymmetry
(the signal before the maximum is asymmetric with respect to the signal after the maximum)
even at $700$ m from the shower core,
and since the SDP is created by the instantaneous deceleration of the 
shower particles at ground level, the asymmetry in
the pulse reflects the asymmetry in the deceleration of ground particles and, therefore,
of the shower disk projection on the ground.

\begin{figure}
\includegraphics[width=0.6\textwidth]{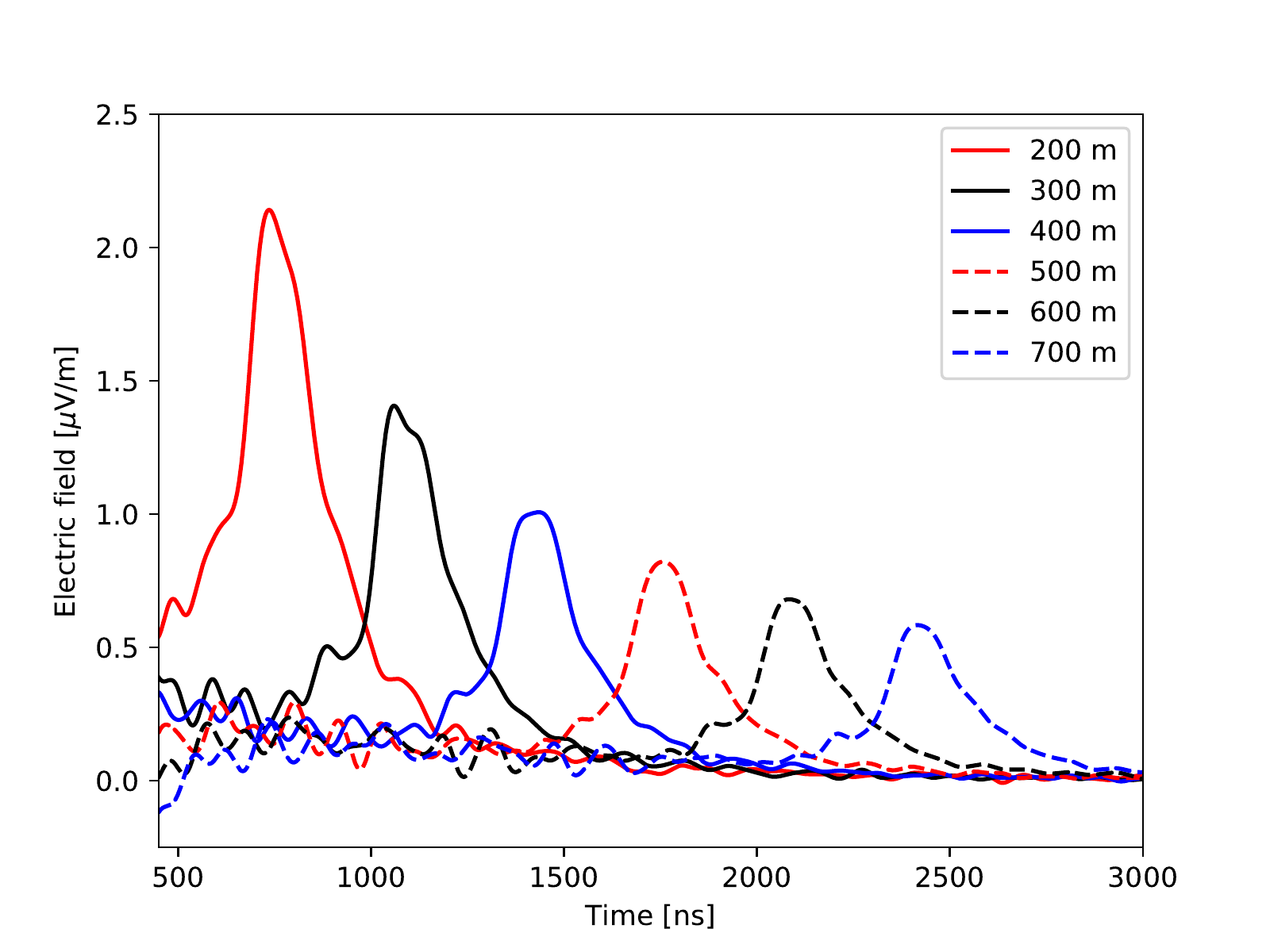}
\caption{SDP electric field (vertical component) as a function of time created by
created by a $1$ EeV proton-induced shower with $30^\circ$ of zenithal angle and
coming from the East ($\phi = 0^\circ$). Observers have been placed north of
the shower core, at regular intervals of $100$ m. The dependence of the arrival
time of the maxima with the distance as well as the amplitude are apparent.
Traces have been filtered with a $10$ MHz sixth-order Butterworth lowpass filter.}
\label{fig:tracesshift}
\end{figure}

The time arrival dependence is verified in Fig.~\ref{fig:timefit}, where we plot the
arrival times (with arbitrary offsets for clarity) 
as a function of distance for showers similar to that in Fig.~\ref{fig:tracesshift}, 
but varying the energy. Each data set has a linear fit superimposed, whose slopes
vary from $3.332$ ns~m${}^{-1}$ to $3.347$ ns~m${}^{-1}$, while the
expected slope is $1/c_n = n/c \approx 3.336$ ns~m${}^{-1}$. The simulated
arrival time is, in fact, in good agreement with the simple model prediction.

\begin{figure}
\includegraphics[width=0.6\textwidth]{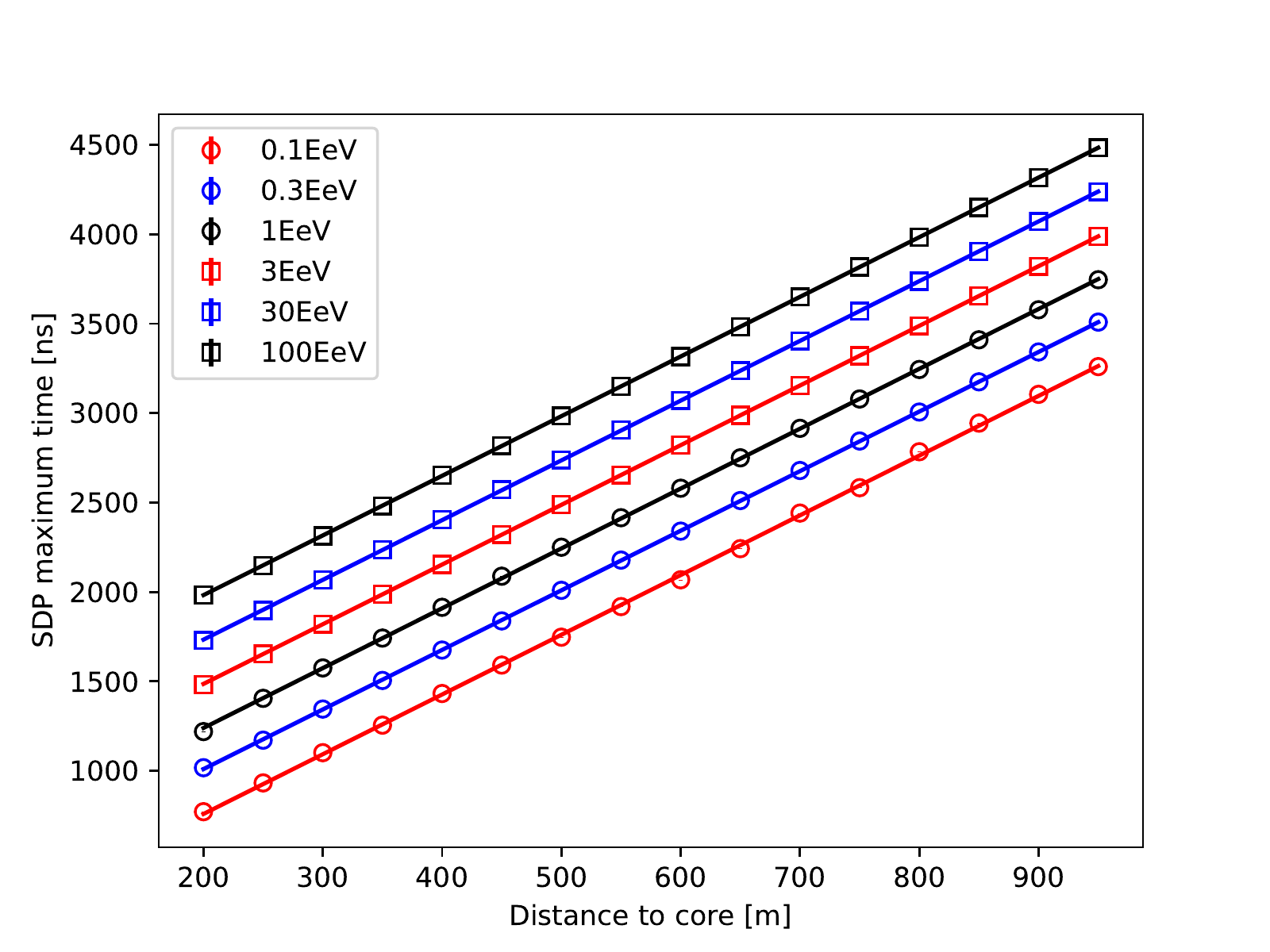}
\caption{Arrival time of the SDP maximum (vertical component) as a function of distance to
the shower core for proton-induced showers of different energies (indicated on the
legend) with $30^\circ$ of zenithal angle and
coming from the East ($\phi = 0^\circ$). Observers have been placed north of
the shower core. The superimposed lines correspond to a linear fit to the data,
with a slope close to the inverse of the speed of light. Times have been offset
for visual clarity. See text for details.}
\label{fig:timefit}
\end{figure}

\subsection{Amplitude of the SDP as a function of the distance to the shower core}
The $1/R$ dependence has been tested in a similar way, but in this case we have
taken the mean of the SDP maximum amplitude for $10$ different simulated showers
with the same geometry and energy as in Fig.~\ref{fig:timefit}. For each set of
amplitudes $A_i$ we have calculated its mean $\langle A \rangle$ and performed
a fit of the form $\log_{10}\langle A \rangle = a + b \log_{10} R $, using as
uncertainty of $\log_{10}\langle A \rangle$ an estimation of the uncertainty of the mean of the logarithm of
the amplitude, that is,
\begin{equation}
\sigma = \sqrt{\frac{1}{n(n-1)} \left[ \langle (\log_{10} A)^2 \rangle - \langle \log_{10} A \rangle^2 \right]}.
\label{eq:logunc}
\end{equation} 
We have plotted the results in Fig.~\ref{fig:sdpfit}. The slope of the fit varies from
$-1.04$ to $1.00$, being compatible with $-1$ (pure radiation field) at $1\sigma$
below $1$ EeV and $2\sigma$ above. In any case, the data are very close to what
one would expect from a pure radiation field, agreeing with our simple model.

\begin{figure}
\includegraphics[width=0.6\textwidth]{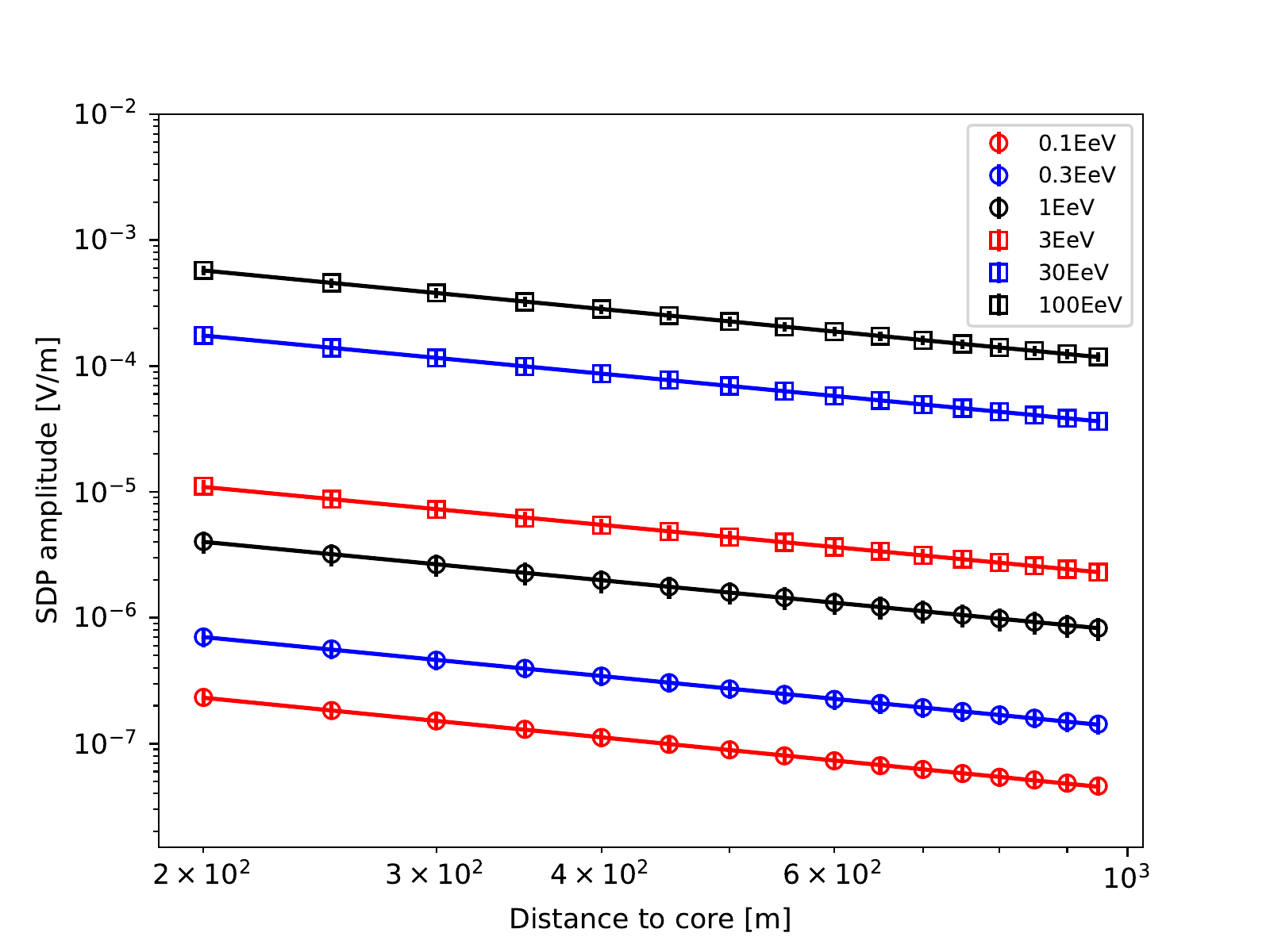}
\caption{Mean amplitude of the SDP maximum (vertical polarization) as a function of distance to
the shower core, averaged over 10 proton-induced showers of different energies (indicated on the
legend) with $30^\circ$ of zenithal angle and
coming from the East ($\phi = 0^\circ$). Observers have been placed north of
the shower core. The superimposed lines correspond to a linear fit to the logarithm of
the amplitude versus the logarithm of the distance, with the uncertainties given
by Eq.~\eqref{eq:logunc} and represented by the small vertical lines at the center
of each marker. The data are compatible with a $1/R$ dependence. See text for details.}
\label{fig:sdpfit}
\end{figure}

\subsection{Amplitude of the SDP as a function of the primary energy}
Another feature of the SDP that can be drawn from Eq.~\eqref{eq:sdp} is that
we expect an amplitude proportional to the number of particles that reach the ground.
This implies that, statistically, the higher the energy, the larger the amplitude of the SDP, although
we do not expect a linear dependence, since the number of particles that arrive to the
ground is not linear with the energy due to the variation of the shower maximum depth
with the primary energy. We can, nonetheless, plot the expected SDP amplitude as a function of
the primary energy for several distances to the shower core, as in Fig.~\ref{fig:sdpenergy},
and fit the logarithm of the SDP amplitude.
The values for the slope are not compatible with $1$, as expected, but the $\chi^2/\textrm{ndof}$ of
the fits are, from smaller to larger distance, $0.75$, $1.07$, $1.57$ and $0.53$, while
the $90\%$ CL for a $\chi^2$ with $4$ degrees of freedom is 
$\chi^2/\textrm{ndof}_{0.9} = 1.99475$, 
meaning that the amplitude can be related to a power of the primary energy for
these particular cases. 
200 m away from the shower core, we find $\propto E^{1.13\pm 0.02}$ 
and 800 m away we find $\propto E^{0.89\pm 0.02}$. However, according to our model, both
should present the same exponent, but this discrepancy is understandable because
the SDP pulse cannot be disentangled from the Askaryan emission coming from the
end of the shower, a feature we will use in Section~\ref{sec:polar} to explain the
polarization of the SDP.

\begin{figure}
\includegraphics[width=0.6\textwidth]{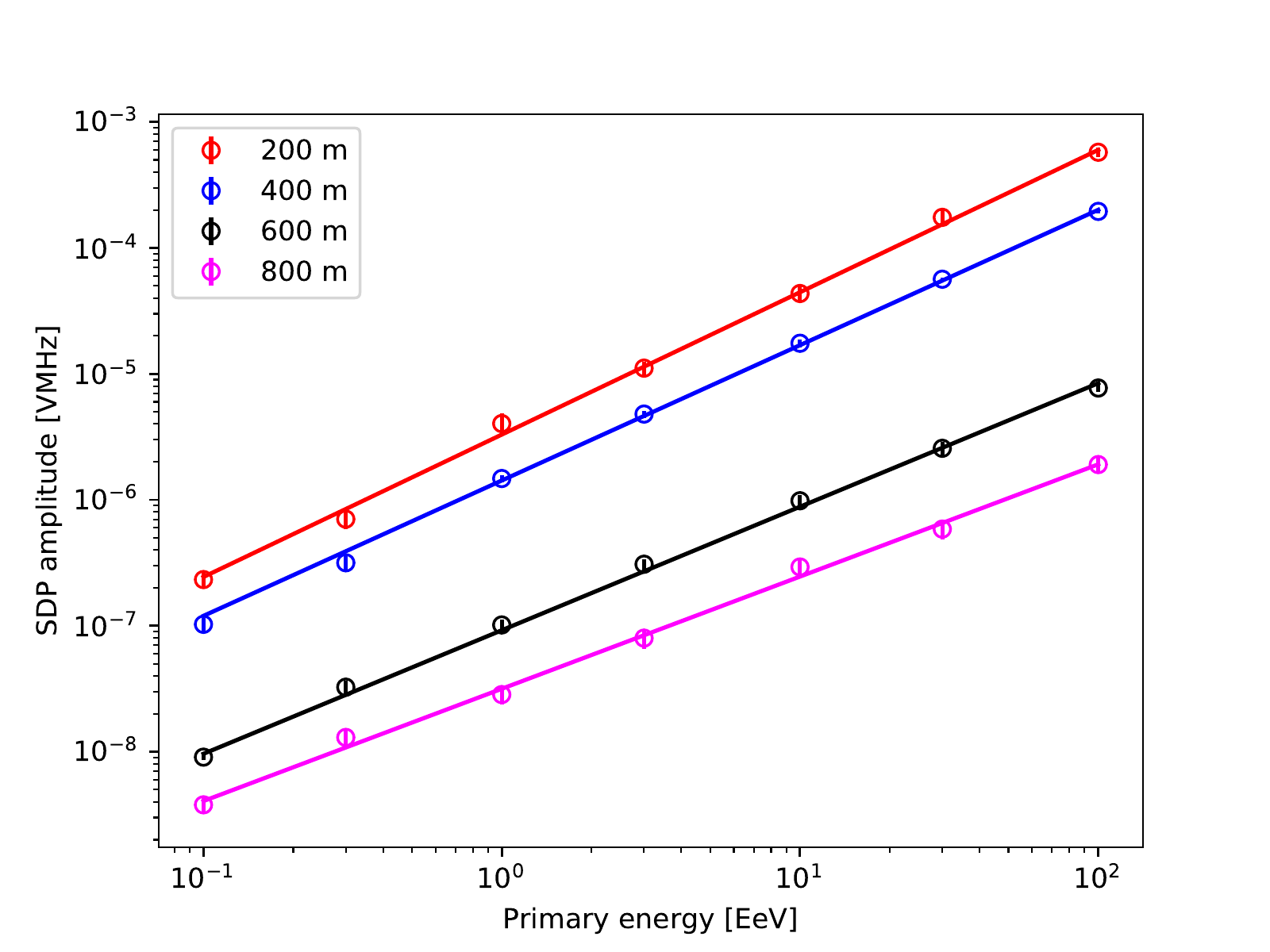}
\caption{Mean amplitude of the SDP maximum (vertical polarization) as a function of 
the primary energy, averaged over 10 proton-induced showers with $30^\circ$ of zenithal angle and
coming from the East ($\phi = 0^\circ$). Observers have been placed north of
the shower core (see legend). The superimposed lines correspond to a linear fit to the logarithm of
the amplitude versus the logarithm of the primary energy, with the uncertainty of the
mean represented by the small vertical lines at the center of each marker.
The simulated data are compatible with a power law of the type 
$A_\textrm{SDP}=a E_\textrm{primary}^b$. See text for details.}
\label{fig:sdpenergy}
\end{figure}

\subsection{Spectrum of the SDP}
In Fig.~\ref{fig:sdpspec} (top) we find an example of the spectrum for the principal pulse
and the SDP created by a $1$ EeV and $30^\circ$ proton-induced shower. The principal pulse
presents the well understood shape of growing electric field up to a certain frequency upon
which the emission ceases to be coherent, falls of and then stabilizes at high frequencies
as an incoherent spectrum. The SDP, on the contrary, is much larger at low frequencies
and becomes incoherent at $\sim 10$ MHz, indicating that the SDP should be observable
only at frequencies below $10$ MHz. Increasing the energy increases the principal pulse,
but also the SDP, each one in a different way. In fact, for a $100$ EeV shower (Fig.~\ref{fig:sdpspec} bottom),
the SDP pulse at low frequencies ($< 2$ MHz) is larger than the principal pulse.

\begin{figure}
\includegraphics[width=0.6\textwidth]{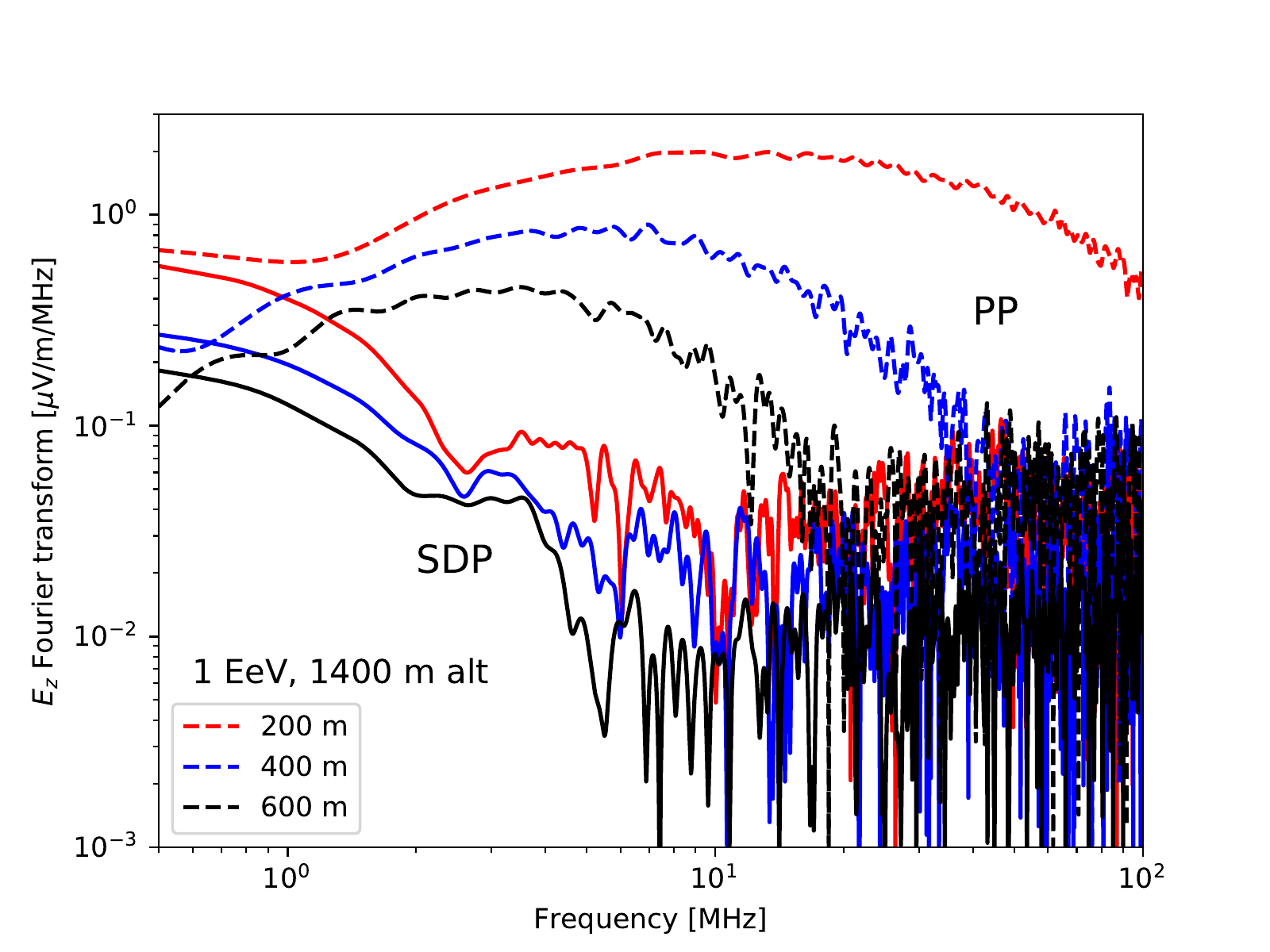}
\includegraphics[width=0.6\textwidth]{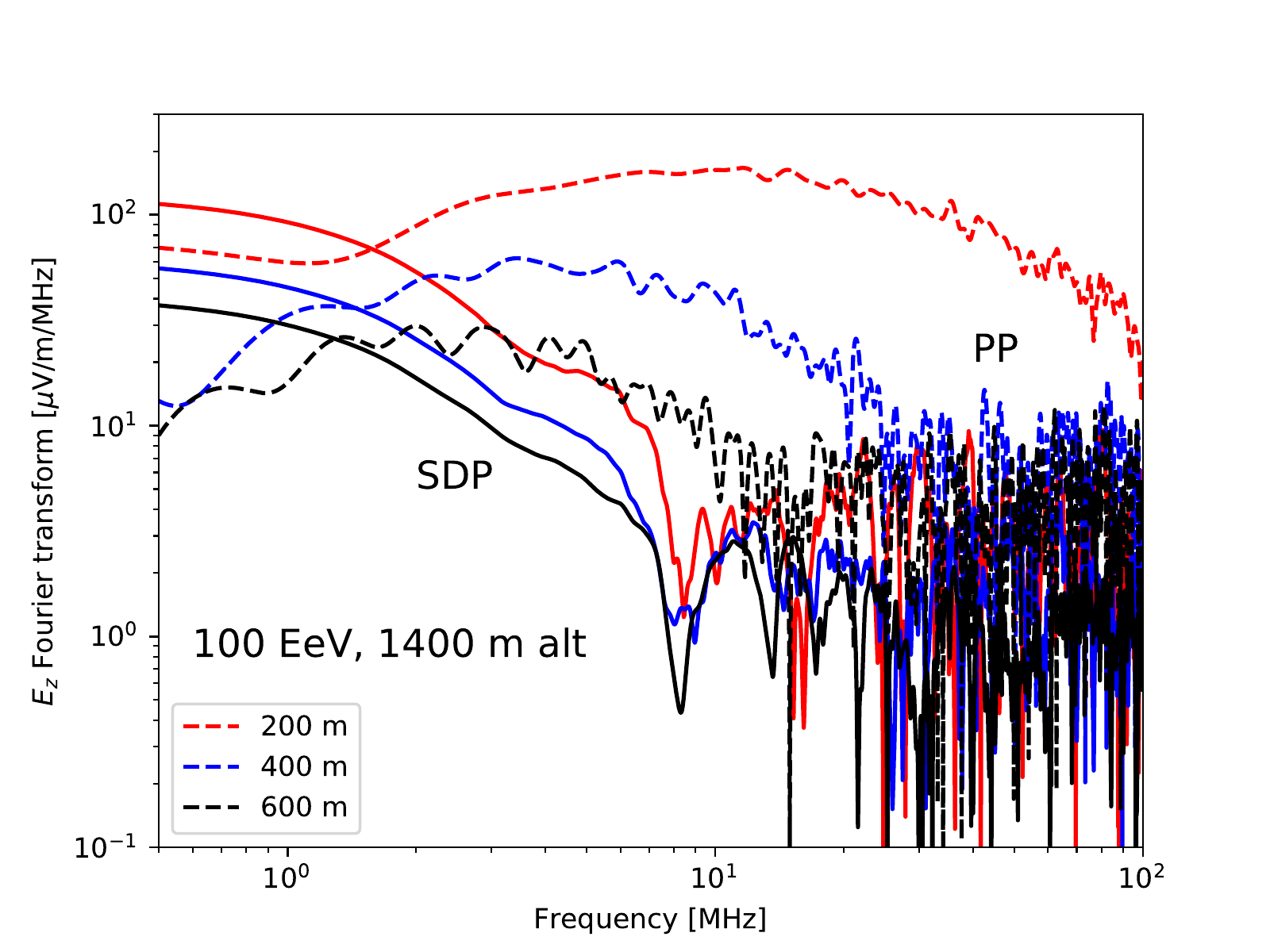}
\caption{Top: Amplitude of the vertical component of the 
Fourier transform of the principal pulse (dashed lines) and the
SDP (solid lines) as a function of frequency
created by a $1$ EeV proton-induced shower with $30^\circ$ of zenithal angle and
coming from the East ($\phi = 0^\circ$). We have placed observers at $200$, $400$
and {$600 \ \textrm{m}$} north of the shower core, located on a ground at $1400$ m
of altitude. Each pulse presents a different frequency
dependence. Bottom: Same as top, but for $100$ EeV of primary energy.}
\label{fig:sdpspec}
\end{figure}

\subsection{Polarization of the SDP}
\label{sec:polar}

We have established that according to Eq.~\eqref{eq:sdp}, we expect a SDP polarization
along the $\hatR \times (\hatR \times \hat{\mathbf{v}})$ direction, where the $\hat{\mathbf{v}}$ direction
is the mean velocity of the particles arriving to the ground divided by its module.
As a first order approximation, 
we can consider $\hat{\mathbf{v}}$ equal to the shower axis unit vector.
This is a good first-order approximation, but we know that there are two important
factors that change the mean velocity, namely:
\begin{itemize}
\item Inclined showers present a ground asymmetry. Particles on the \emph{early} region of
the ground, closer to the shower axis, will present a different velocity distribution with
respect to the particles on the \emph{late} region.
\item The geomagnetic field changes the velocity of the particles. Positively charged particles
will be deflected in one direction and the negatively charged particles in the opposite direction.
Since negative particles are more predominant, due to the Askaryan effect, this will
induce a total mean velocity different from the shower axis.
\end{itemize}
This means that approximating the mean velocity as the shower axis should be reliable
for vertical showers and with negligible geomagnetic effects. For inclined showers or
strong magnetic fields, we cannot expect it to be a good approximation.

One thing to consider is that the SDP is created by an excess of negative particles
that comprise the queue of the air shower,
meaning it is irremediably intertwined with the Askaryan effect. In other words, whenever
we find a physical SDP, an electric field created by the excess charge should be present
and it cannot be separated. With this in mind it is possible to improve the model with
the presence of an Askaryan field, with a polarization pointing towards the shower axis,
which we will call $\hat{\mathbf{r}}_A$.
Let us assume that the SDP has a proportionality constant $\alpha_S$, while the
Askaryan field has a proportionality constant $\alpha_A$, and let us transform
Eq.~\eqref{eq:sdp} accordingly.
\begin{equation}
\mathbf{E}_{\textrm{SDP,mean}} =
\alpha_S \left[ \hatR \times (\hatR \times \hat{\mathbf{v}}_\textrm{axis}) \right]
+ \alpha_A\hat{\mathbf{r}}_A
\label{eq:sdpa}
\end{equation}
Eq.~\eqref{eq:sdpa} represents the mean SDP field, averaged over the time the SDP is
seen by the observers. The coefficients $\alpha_S$ and $\alpha_A$ can be positive or
negative, in order to give the correct polarization sign. It is worth noting that these two
coefficients are a function of the shower zenith angle and of the distance to the
shower core (in the case of $\alpha_S$) and shower axis (in the case of
$\alpha_A$), since that will influence the number
of particles arriving to the ground and also the amount of excess charge when they arrive.
These dependences,
$\alpha_{S} \approx f_{S}(\theta)/R$ and
$\alpha_{A} \approx f_{A}(\theta)/g(\theta,R_\textrm{axis})$,
where $R_\textrm{axis}$ is the distance to the shower axis and $R$ is the
distance to the shower core,
set the limits of what we can do with this
simple model.

Let us assume a $1$ EeV shower coming from the East and let us place observers
on a $500 \ \textrm{m}$ radius ring with the shower core as the center. 
The ground is located at $1400$ m and a
$47.57\ \mu$T magnetic field having an inclination of $I = 62.94^\circ$ and a
declination of $D = 0.42^\circ$. The shower unit vector can be written as:
\begin{equation}
\hat{\mathbf{v}}_\textrm{axis} = -\sin\theta\, \hat{x} - \cos\theta\,\hat{z},
\label{eq:pim}
\end{equation}
while the unit core-observer vector can be written with the help of the observer's
azimuth:
\begin{equation}
\hatR = \cos\phi  \,\hat{x} + \sin\phi\,\hat{y}.
\label{eq:pam}
\end{equation} 
The Askaryan field vector is written for this case as:
\begin{equation}
\hat{\mathbf{r}}_A = -\cos\phi\cos\theta\,\hat{x}
-\sin\phi\,\hat{y}
+ \cos\phi\sin\theta\,\hat{z}
\label{eq:pum}
\end{equation}
Putting Eqs.~\eqref{eq:pim}, \eqref{eq:pam} and~\eqref{eq:pum} in Eq.~\eqref{eq:sdpa},
we present the East-West, North-South and Vertical mean electric fields predicted for
a shower coming from the east.
\begin{equation}
\langle \mathbf{E}_{SDP,EW} \rangle = \alpha_S \sin^2\phi\sin\theta - \alpha_A\cos\phi\cos\theta
\label{eq:sdpew}
\end{equation}
\begin{equation}
\langle \mathbf{E}_{SDP,NS} \rangle= -\alpha_S\sin\phi\cos\phi\sin\theta - \alpha_A\sin\phi
\label{eq:sdpns}
\end{equation}
\begin{equation}
\langle \mathbf{E}_{SDP,V} \rangle = \alpha_S\cos\theta + \alpha_A\cos\phi\sin\theta
\label{eq:sdpv}
\end{equation}
After fixing a zenithal angle, these three equations allow us to fit the polarization as a
function of the observer's azimuth $\phi$, leaving $\alpha_S$, $\alpha_A$ as free
parameters. We show in Fig.~\ref{fig:sdppol} the results of the simulations and the fits
using Eqs.~\eqref{eq:sdpew}, \eqref{eq:sdpns} and \eqref{eq:sdpv}. Results are remarkably
good for low zenith angle showers, while with increasing angle the quality of the fit
decreases. The shifting of the extrema is an effect caused by the inclination
of the shower, having checked that it is not due to the geomagnetic field (we checked
with simulations performed with a null magnetic field). The geomagnetic field, however
is responsible for the different height of the maxima in the EW polarization, something
that is expected given it is fixed along the EW direction. We could improve this model,
since it is known that the inclination of the shower will produce an Askaryan field on
the ground with an elliptical symmetry \cite{washington}, but this is outside of the
scope of this paper. Our goal was to demonstrate that a simple model can be
useful to understand qualitatively the main characteristics of the SDP polarization.

For showers above $40^\circ$, the polarization becomes more complicated, whereas
for showers above $55^\circ$ the SDP pulse becomes quite feeble and subject to
huge fluctuations.

\begin{figure}
\includegraphics[width=0.49\textwidth]{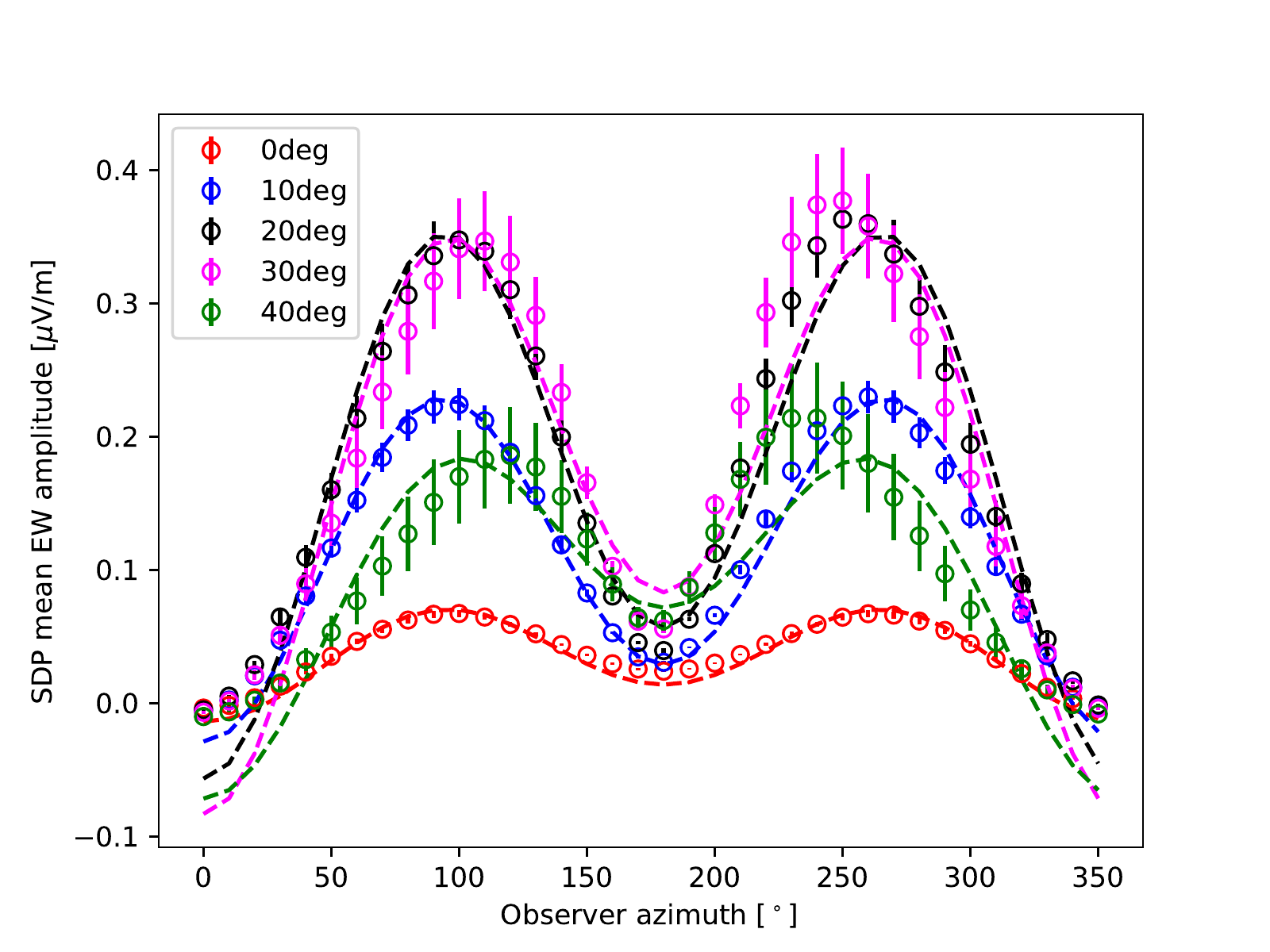}
\includegraphics[width=0.49\textwidth]{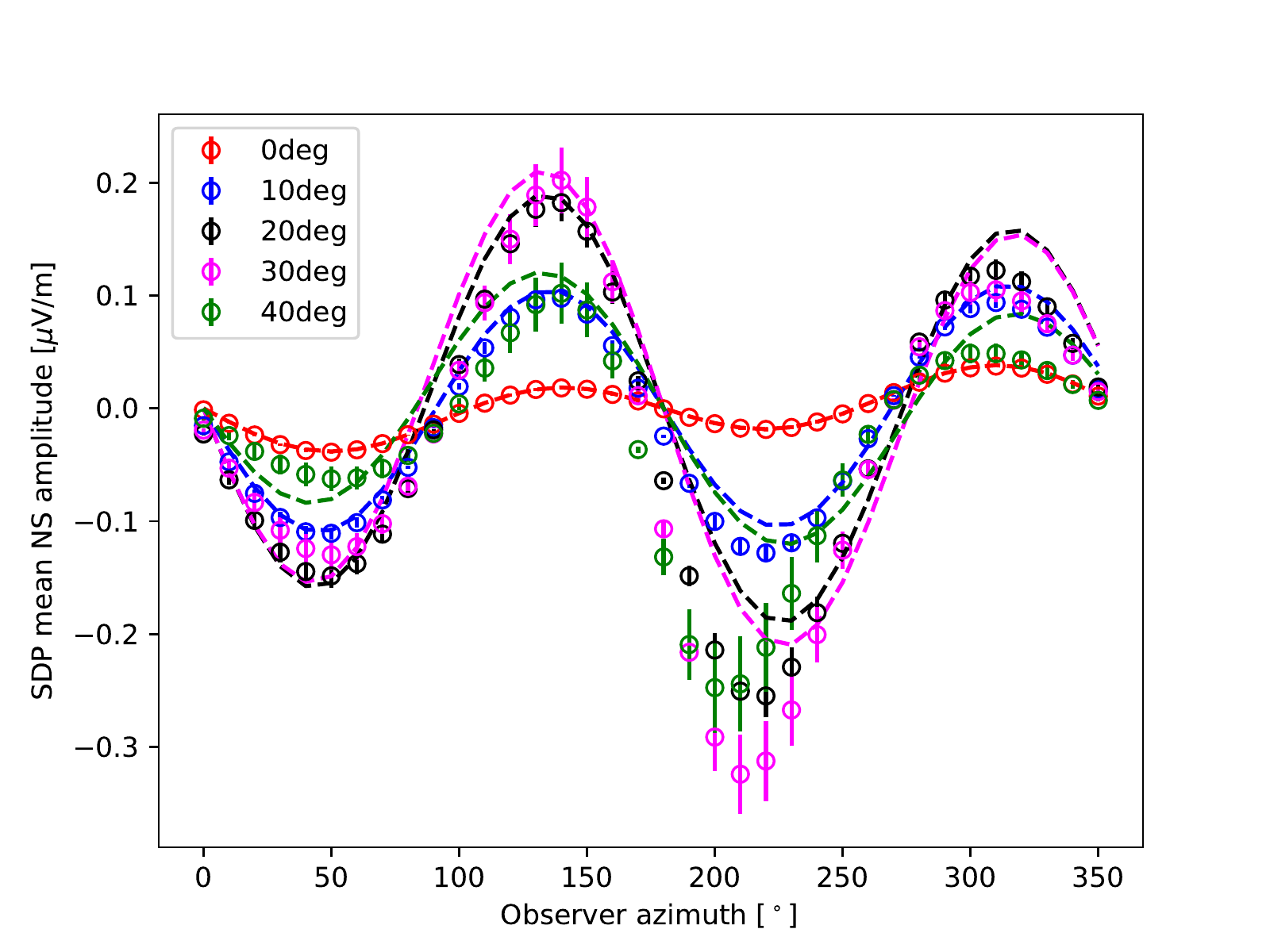}
\includegraphics[width=0.49\textwidth]{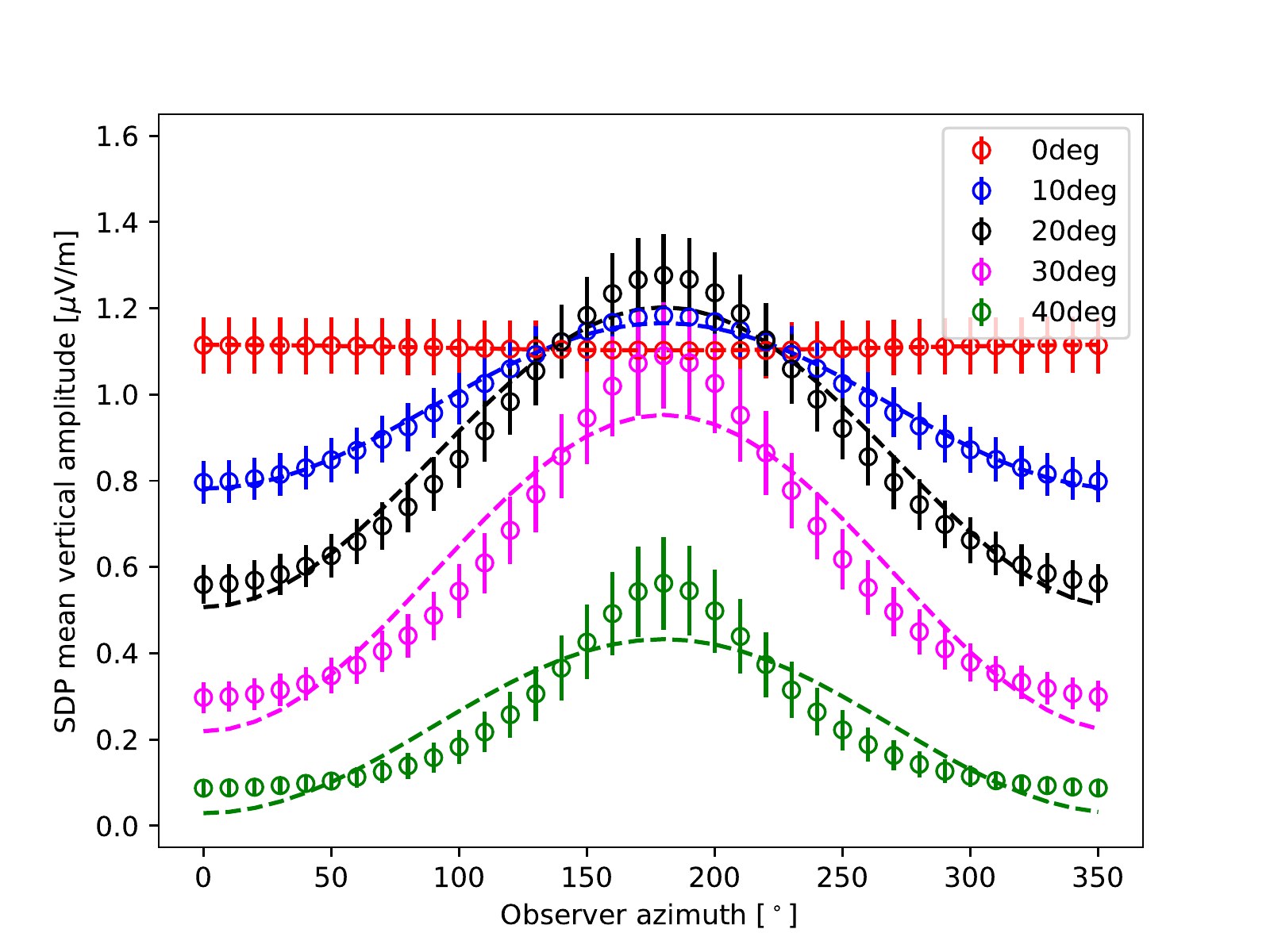}
\caption{Mean value of the SDP electric field (as a function of the observer's azimuthal angle) 
created by several proton-induced $1$ EeV
showers coming from the east with different zenith angles indicated by the legend. The
field has been averaged over a 500 ns window. Observers have been placed in a 500 m
radius ring around the shower core. The points represent the simulation data and the
dashed lines are the fit results according to Eqs.~\eqref{eq:sdpew}, \eqref{eq:sdpns}
and \eqref{eq:sdpv}. Top left: EW polarization. Top right: NS polarization. Bottom:
vertical polarization.}
\label{fig:sdppol}
\end{figure}

\subsection{SDP amplitude \emph{vs} number of ground particles}
According to Eq.~\eqref{eq:sdp}, we expect the maximum amplitude of the SDP
to be proportional to the excess of negatively charged 
particles arriving at the ground. If we choose
the vector $\hatR$ to be perpendicular to $\mathbf{v}$, the total maximum
amplitude of the pulse in our simple model is equal to
\begin{equation}
|\mathbf{E}_{\mathrm{SDP,max}}\xt| =
\frac{Nq}{4\pi\epsilon R} \delta\left( t - \frac{nR}{c} \right) \propto N,
\end{equation}
which is effectively proportional to the number of ground particles. Let us place
an observer 200 m north of the shower core and let us simulate the electric
fields from showers coming from the East at zenith angles from $0$ to
$80^\circ$ and primary energies ranging from $0.1$ EeV to $100$ EeV.
We show the results in Fig.~\ref{fig:npart} for a ground at $180$ m of
altitude, similar to that the EXTASIS experiment rests on. We show in
Fig.~\ref{fig:npart}, top left, the number of electrons and positrons arriving
at the ground as a function of the primary energy and shower zenith angle,
while at the top right part of the figure we find the same plot for the maximum
total amplitude of the SDP. Both figures are similar, indicating a correlation
between number of particles and pulse amplitude. We show in
Fig.~\ref{fig:npart}, bottom left, the same plot for the vertical polarization
and at the bottom right the EW polarization.

\begin{figure}
\includegraphics[width=0.49\textwidth]{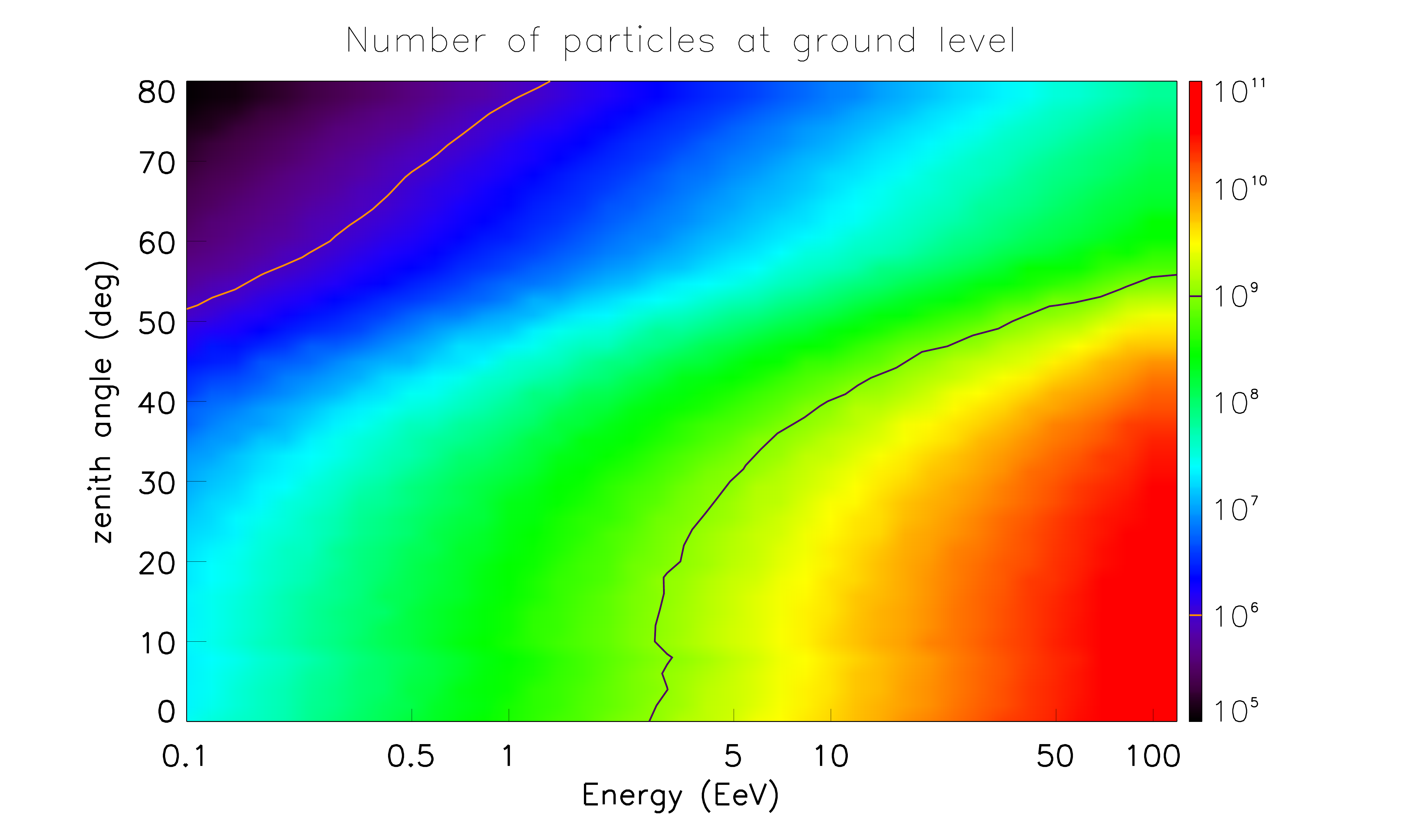}
\includegraphics[width=0.49\textwidth]{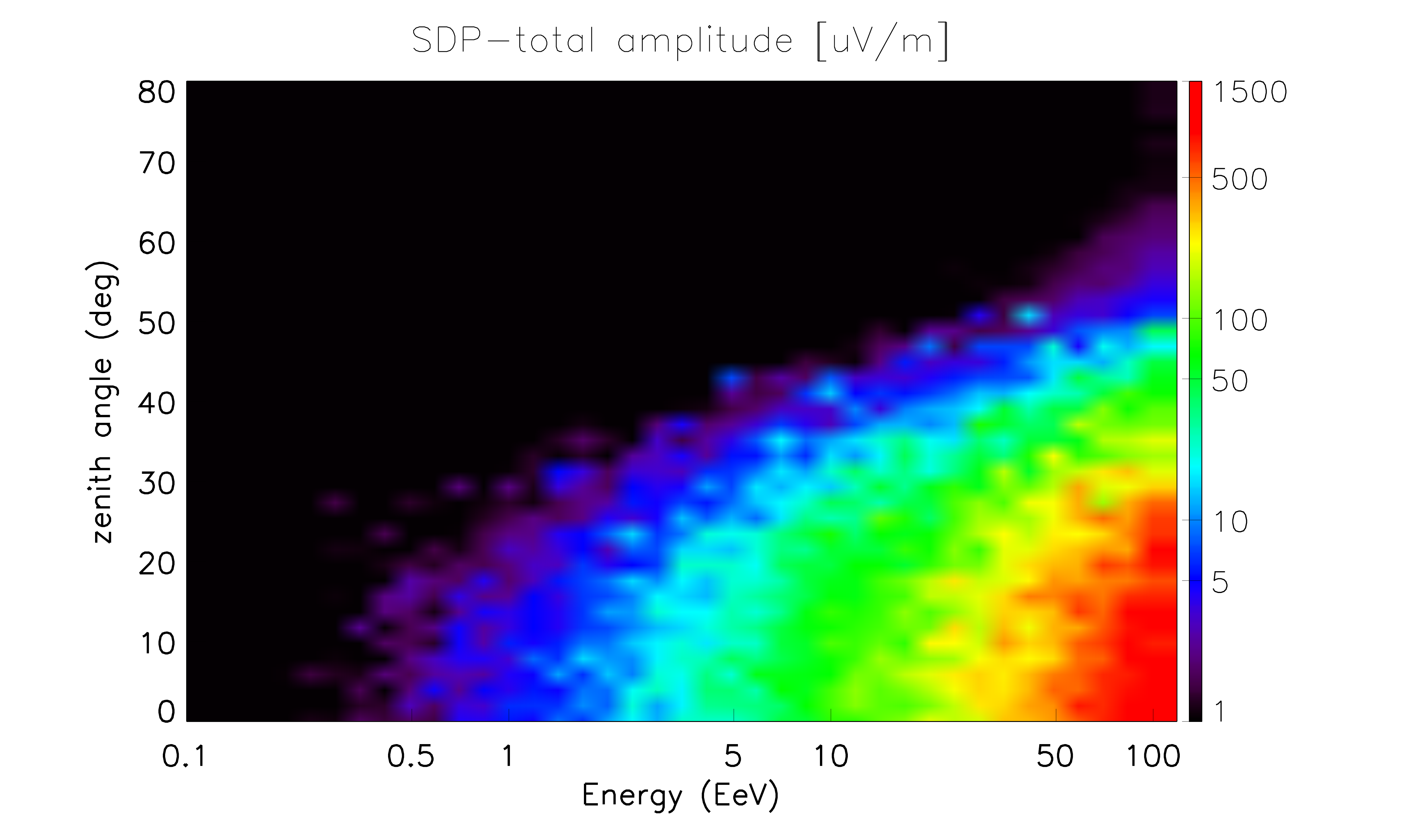}
\includegraphics[width=0.49\textwidth]{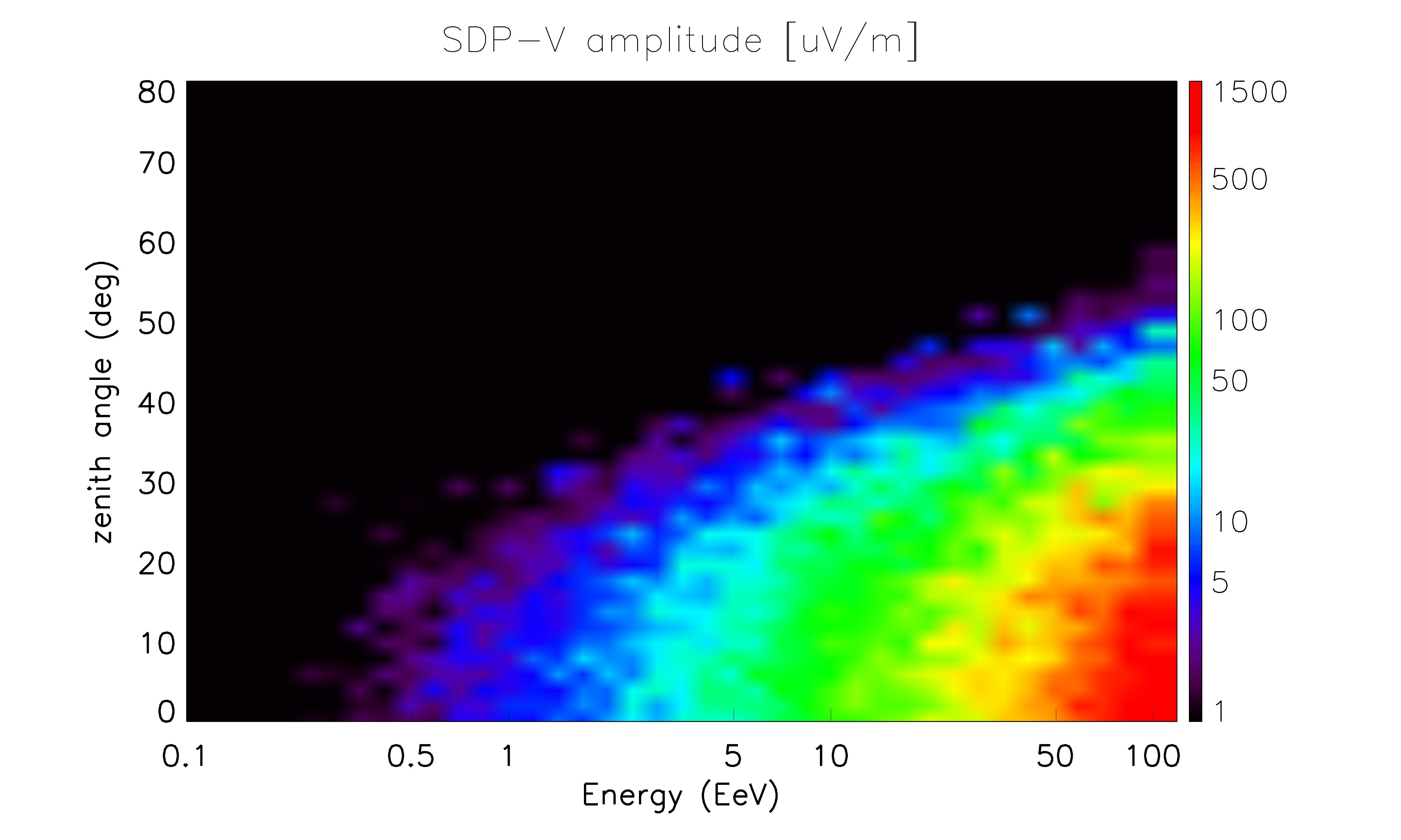}
\includegraphics[width=0.49\textwidth]{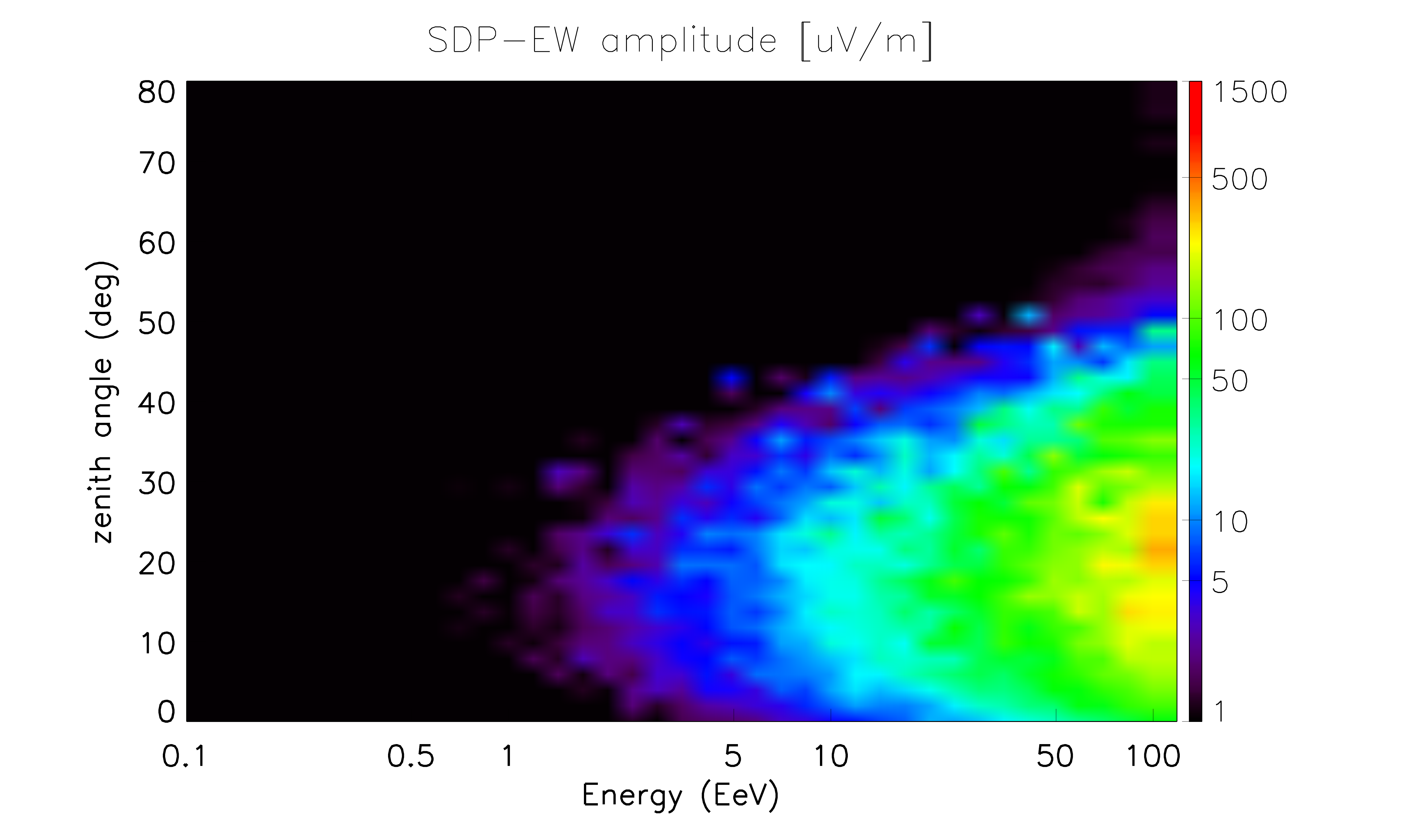}
\caption{Top left: two-dimensional color plot of the number of electrons
and positrons arriving at the ground (180 m of altitude) as a function of 
the primary proton energy
and shower zenith angle. Each bin contains the mean number of particles
averaged over 100 showers. Top right: Two-dimensional color plot of the
total SDP maximum amplitude (three polarizations) as a function of the
primary proton energy and shower zenith angle. Each bin contains the mean
amplitude averaged over 5 showers.
Bottom left: Same as top right, for the vertical polarization.
Bottom right: Same as top right, for the EW polarization.}
\label{fig:npart}
\end{figure}

Let us increase the altitude to $1400$ m. An increase in the number of particles
and the SDP amplitude is expected, which is precisely what we find in 
Fig.~\ref{fig:npart1400}. Supposing a detection threshold of 5 $\mu$V/m,
detectable showers at $180$ m of altitude should have an energy larger than
$1$ EeV (see the region at the right of the 5 $\mu$V/m contour line, 
Fig.~\ref{fig:npart}, top right). For the EXTASIS
site and the projected livetime of $\sim 2,3$ years we do not expect 
showers having more energy than $10$ EeV,
so we aim to detect the SDP coming from 
showers with energies $> 1$ EeV and having a zenith
angle $< 30^\circ$. The more vertical the shower, the easier it should be to
detect. For a site at $1400$ m of altitude (Fig.~\ref{fig:npart1400}, right),
the range of detectability is larger, and it is even larger for a $2650$ m altitude,
similar to the one at the GRAND site \cite{grand}.

\begin{figure}
\includegraphics[width=0.49\textwidth]{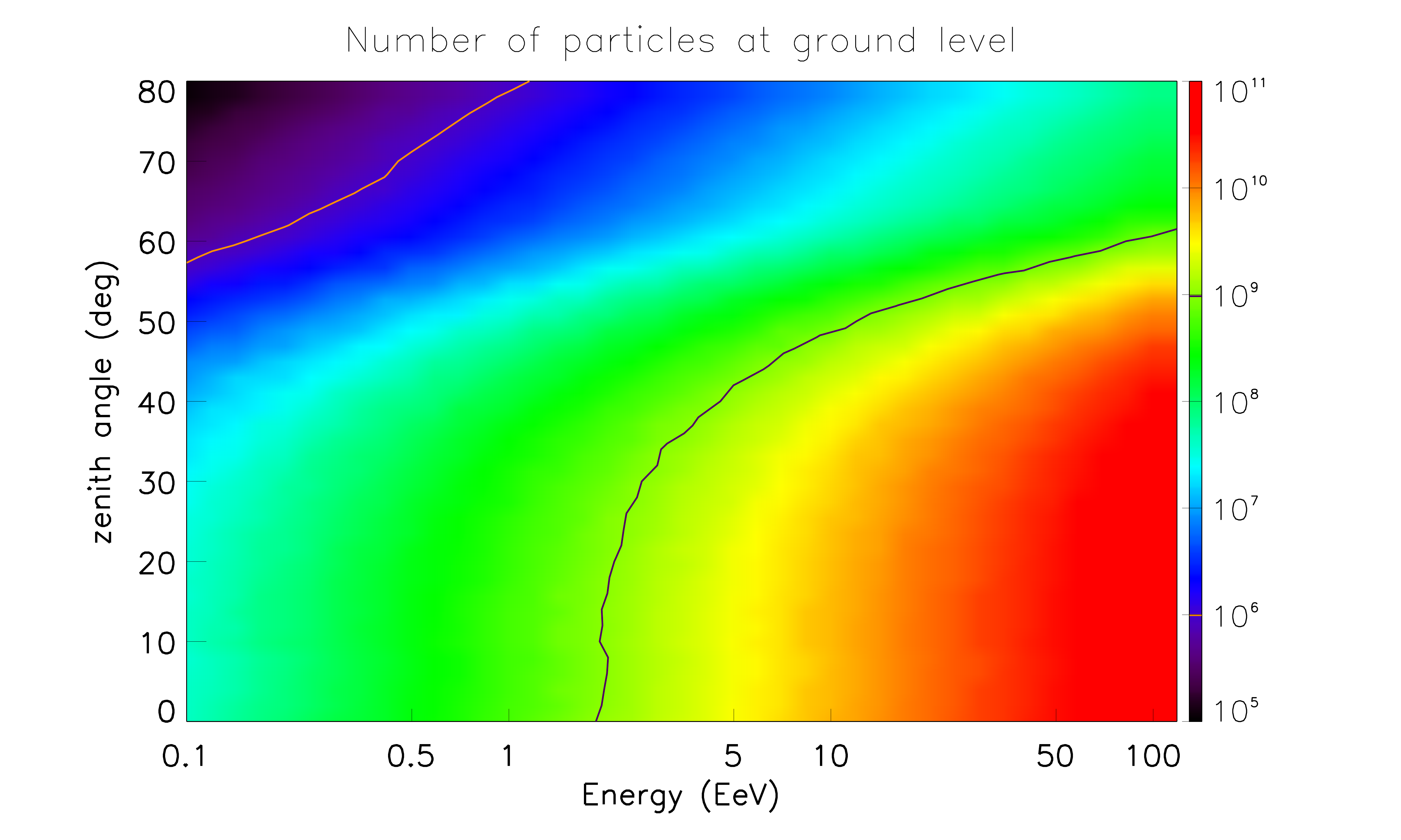}
\includegraphics[width=0.49\textwidth]{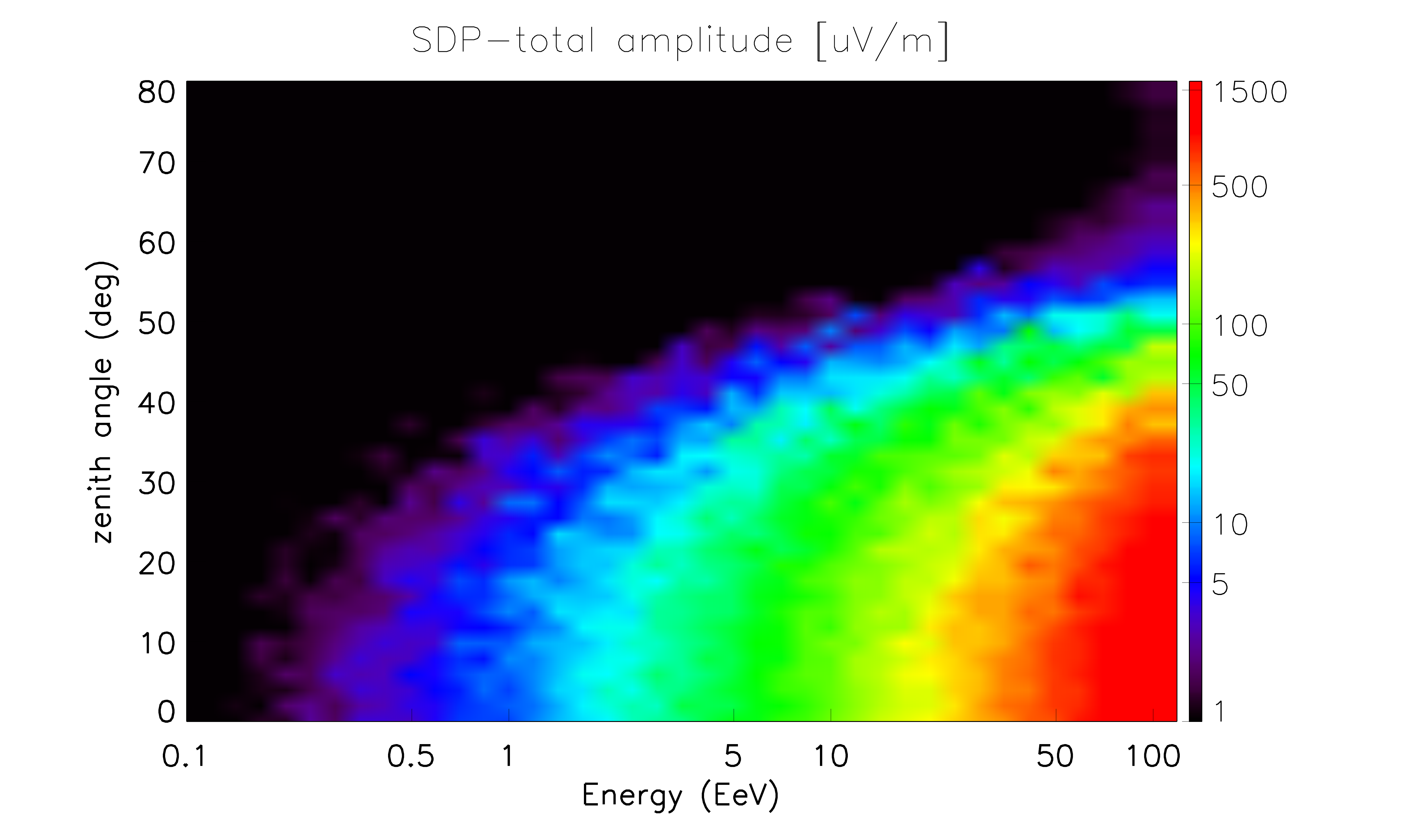}
\caption{Same as Fig.~\ref{fig:npart}, top, but with a ground altitude equal
to 1400 m.}
\label{fig:npart1400}
\end{figure}

\section{Summary and conclusions}

We have presented a time-domain equation of the electric field produced by
a particle track, defined as a charged particle that accelerates instantaneously,
moves at a constant speed and instantaneously decelerates. Special emphasis
has been put on the conservation of the total charge so as to ensure that the
equation is a valid solution of Maxwell's equations and therefore gives the
correct physical field. Explicitely conserving the charge results in the apparition
of two impulse fields coming from the first and last points of the trajectory of
the particle track and that needs to be added to the total field. We have checked
that our complete equation reduces in the far-field limit ($kR \gg 1$) to the
ZHS formula, which has been proven to be a correct expression for the far-field
emission of a particle track.

We have also solved the exact electric field for a track in frequency (Fourier) domain,
starting from the same physical problem (same set of charge densities and
current densities) but solving Maxwell's equations in frequency domain.
We have shown that the result is analytically equivalent to the Fourier transform
of the time-domain electric field. This is a check of the correctness of the formula
as well as a confirmation that working with the time-domain formula and then
transforming to frequency domain is mathematically sound, since
both equations are Fourier transforms of each other. Limitations
in validity and precision come from the numerical step chosen for the evaluation
of the time-domain field and from the discrete Fourier transform employed.
Moreover, the frequency-domain version of the electric field for a track is
numerically equivalent to another frequency-domain formula \cite{prd87} obtained
with the same hypotheses but with a different way of calculating the scalar
potential, in what constitutes another check for our expression.

We have implemented the formula in the SELFAS Monte Carlo code so as to
know the emission it predicts for EAS. In doing so, we neglected the static
fields created after the particle flight since this physical situation is not going
to be given in a real EAS, for the atmosphere will tend to regain electrostatic
equilibrium as soon as possible. After the implementation, the results of our
simulations show that for a frequency larger than 20 MHz and horizontal polarizations
(the case for experiments like AERA, CODALEMA, LOFAR or Tunka-Rex), the
far-field approximation can be used with an error of the order of $\sim 1\%$ or lower
at distances greater than $100$ m from the shower core.
However, for smaller distances to the shower core one may consider using
the exact formula for a particle track. In the case of measuring at a frequency
lower than $20$ MHz or measuring the vertical compontent of an EAS, 
as it is the case for the EXTASIS experiment, the use of the exact formula
(Eq.~\eqref{eq:Etime2}) is unavoidable if one is to know the accurate electric field.
For experiments conducted in dense media, such as ARA, we expect the far-field
approximation to be valid almost always due to the much smaller size of the
shower, as shown in \cite{prd87}.

We have shown a low-frequency impulse field created by the coherent
deceleration of the shower front upon its arrival to the ground, whom we have
called the Sudden Death Pulse (SDP). This pulse vanishes for frequencies above
$20$ MHz, which means that such a field could be detectable with dedicated
low-frequency experiments like EXTASIS. In a first order approximation, we have
calculated only the direct emission of the decelerating shower front, 
ignoring the induced surface wave.

We have studied the properties of the direct emission of the SDP. The arrival time
of the shower maximum is linear with the distance from the observer to
the shower core, and the proportionality constant is compatible with the inverse
of the speed of light. The amplitude of the SDP decays with the inverse of the
distance to the shower core as well, which means that the SDP behaves similarly
to a radiation field. These results can be explained using a simple model consisting
in a long track containing all the excess of charged particles from the shower that
arrive to the ground and calculating the field created in the instant the track
suddenly decelerates. This model also predicts a polarization
as a function of the observer's azimuth angle that, coupled with
a simplified Askaryan polarization, gives a good quantitative agreement for
vertical showers and explains qualitatively the polarization for more inclined showers.

The amplitude of the SDP has been found to be a monotonously increasing function
of the primary energy, and in particular it is proportional to the number of particles
arriving to the ground. The vertical polarization is favored for vertical showers while
the horizontal polarizations are more present the more inclined is the shower, as
suggested by the simple model proposed here.

One important \emph{caveat} to keep in mind is that while 
the exact field of a particle track is needed if we need to calculate the near-field
emission of an EAS, one has to know the response of the measuring antenna
to an incoming near field emission to elucidate the final antenna voltage. 
In this work we have been concerned with
obtaining a physical electric field, but the link to the antenna voltage, which is
what is measured eventually, lies outside the scope of this paper.

Finally, while the SDP explains some of the properties of the pulses measured
by the low-frequency experiments in the past, the SDP amplitudes obtained in the
present work are not large enough to explain all the experimental data.
A study on the influence of the surface wave may help us to solve this discrepancy.
This work is underway.

\bibliography{formula_paper}

\end{document}